\documentclass{article}
\usepackage{arxiv}
\usepackage[utf8]{inputenc} 
\usepackage[T1]{fontenc}    
\usepackage{hyperref}       
\usepackage{url}            
\usepackage{booktabs}       
\usepackage{amsfonts}       
\usepackage{nicefrac}       
\usepackage{microtype}      
\usepackage{lipsum}		
\usepackage{graphicx}
\usepackage{multirow}
\usepackage{amsmath}
\usepackage{subfigure} 
\usepackage{chapterbib}	
\usepackage[square,numbers,sectionbib]{natbib} 
\usepackage{appendix}   
\title{Determining the gravity potential with the CVSTT technique using two hydrogen clocks}

\author{ \href{https://orcid.org/0000-0001-6505-2870}{\includegraphics[scale=0.06]{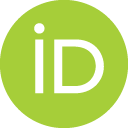}\hspace{1mm}Kuangchao~Wu} \\
	School of Geodesy and Geomatics\\
	Wuhan University\\
	Wuhan 430079, China \\
	\texttt{Kcwu@whu.edu.cn} \\
	\And
	\href{https://orcid.org/0000-0002-9267-5982}{\includegraphics[scale=0.06]{orcid.png}\hspace{1mm}Wen-Bin~Shen}\thanks{Corresponding author, W B Shen,Time and Frequency Geodesy Center, School of Geodesy and Geomatics/Key Laboratory of Geospace Environment and Geodesy of Ministry of Education/State Key Laboratory of Information Engineering in Surveying, Mapping and Remote Sensing, Wuhan University, Wuhan 430079, China} \\
	School of Geodesy and Geomatics\\
State Key Laboratory of Information Engineering in Surveying\\
	Wuhan University\\
	Wuhan 430079, China\\
	\texttt{wbshen@sgg.whu.edu.cn} \\
	\And
	\href{https://orcid.org/0000-0003-4349-0460}{\includegraphics[scale=0.06]{orcid.png}\hspace{1mm}Xiao~Sun} \\
State Key Laboratory of Information Engineering in Surveying\\
	 Wuhan University\\
	Wuhan 430079, China\\
	\texttt{xsun@whu.edu.cn} \\
	\And
	\href{https://orcid.org/0000-0003-3437-4820}{\includegraphics[scale=0.06]{orcid.png}\hspace{1mm}Chenghui~Cai} \\
	School of Geodesy and Geomatics\\
	Wuhan University\\
	Wuhan 430079, China\\
	\texttt{chcai@whu.edu.cn} \\
	\And
	\href{https://orcid.org/0000-0002-7261-1068}{\includegraphics[scale=0.06]{orcid.png}\hspace{1mm}Ziyu~Shen} \\
	School of Resource and Environment\\
	Hubei University of Science and Technology\\
	Xianning, Hubei, China.\\
	\texttt{theorhythm@foxmail.com} \\
}

\begin{document}
\maketitle
\begin{abstract}
According to general relativity theory (GRT), by comparing the frequencies between two precise clocks at two different stations, the gravity potential (geopotential) difference between the two stations can be determined due to the gravity frequency shift effect. Here, we provide experimental results of geopotential difference determination based on frequency comparisons between two remote hydrogen atomic clocks, with the help of common-view satellite time transfer (CVSTT) technique. For the first time we apply the ensemble empirical mode decomposition (EEMD) technique to the CVSTT observations for effectively determining the geopotential-related signals. Based on the  net frequency shift between the two clocks in two different periods, the geopotential difference between stations of the Beijing 203 Institute Laboratory (BIL) and Luojiashan Time--Frequency Station (LTS) is determined. Comparisons show that the orthometric height (OH) of LTS determined by the clock comparison is deviated from that determined by the Earth gravity model EGM2008 by (38.5$\pm$45.7)~m. The results are consistent with the frequency stabilities of the hydrogen clocks (at the level of $10^{-15}$~day$^{-1}$) used in the experiment. Using more precise atomic or optical clocks, the CVSTT method for geopotential determination could be applied effectively and extensively in geodesy in the future.
\end{abstract}

\section{Introduction}
Precise determination of the Earth's gravity potential (geopotential) field and orthometric heights (OHs) are main tasks in geodesy \cite{Mazurova.2016, Mazurova.2017}. With the rapid development of time and frequency science, high-precision atomic clock manufacturing technology provides an alternative way to precisely determine geopotentials and OHs, which has been extensively discussed in recent years \cite{Shen.2016, Lion.2017, Mcgrew.2018, Tanaka.2021}, opening a new era of time--frequency geoscience \cite{Kopeikin.2018, Mehlstaubler.2018, Shen.2019, Puetzfeld.2019}.

General relativity theory (GRT) states that a precise clock runs at different rates in different places with different geopotentials \cite{Einstein.1915, Lion.2017}. Consequently, the geopotential difference between two arbitrary points can be determined by comparing the frequencies of two precise atomic clocks \cite{Bjerhammar.1985, Shen.1993, Mai.2013, Shen.2011}. In order to determine geopotentials with an accuracy of 0.1~m$^{2}$~s$^{-2}$ (equivalent to 1~cm in OH), clocks with frequency stabilities of  $10^{-18}$ are required.

High-performance clocks have been intensively developed in recent years. Optical atomic clocks (OACs) with stabilities around the $10^{-18}$ level have been successively generated \cite{Campbell.2017, Samuel.2019}, and in the near future mobile high-precision satellite-borne optical clocks will be in practical use \cite{Liu.2018}. This enables 1~cm level geopotential determination and, potentially, unification of the world height system (WHS) is promising \cite{Mcgrew.2018, Mueller.2018, Shen.2019b}.

To realize precise frequency comparisons for geopotential determination between two remote clocks, we need not only clocks with high stabilities, but also a reliable time-frequency transfer technique that can precisely compare the frequencies. As early as the 1980s it was proposed that the common-view satellite time transfer (CVSTT) technique can be used for comparing frequencies between remote clocks \cite{Allan.1985}. This technique is adopted by the International Bureau of Weights and Measures (BIPM) as one of the main methods for transferring international atomic time (TAI) signals \cite{Allan.1994, Defraigne.2015}. Using this method, the uncertainty of comparing remote clocks may reach several nanoseconds \cite{Lewandowski.1999, Ray.2003, Rose.2014}.

There are two kinds of methods for determining geopotentials via clocks. One is to compare the frequency shift of precise clocks via opticla fibers (or coaxial cables) at ground \cite{Bjerhammar.1985, Takano.2016, Shen.2019b}. Another one is to conduct the frequency comparison between clocks at two ground stations via the  satellite's links \cite{Kopeikin.2016, Shen.2017, Cai.2020}. Some successful clock-transportation experiments have been conducted for testing gravitational redshift or determining geopotentials via fiber links \cite{Lisdat.2016, Takano.2016, Grotti.2018, Takamoto.2020}. Grotti and his colleagues made a clock-transportation experiment using OACs with stabilities at the $10^{-17}$ level\cite{Grotti.2018}. By frequency comparison they determined geopotential difference as 10,034(174)~m$^{2}$~s$^{-2}$, which is roughly in agreement with the value of 10,032.1(16)~m$^{2}$~s$^{-2}$ determined independently by geodetic means. However, there are seldom studies on geopotential determination using the satellite time--frequency transfer technique. In fact, the satellite's link approach is very prospective because it is not constrained by geographical conditions, for instance connecting two continents separated by oceans, which is extremely costly using fiber-link method. In 2016, using a transportable hydrogen atomic clock with stability of $7\times10^{-15}$~day$^{-1}$, Kopeikin and his colleagues provided experimental results of geopotential difference determination via the satellite's link approach \cite{Kopeikin.2016}. Their experimental result of OH difference is ($725\pm64$)~m,  which has a discrepancy as large as about 133~m compared to the value ($858.9\pm0.1$)~m obtained by conventional method, which is not consistent well with the corresponding accuracy of 64~m. The large deviation might be due to their comparison duration being quite short, which lasts about 24~hr in total. 

Here we focus on frequency comparisons of clocks via satellite's link, and determine the geopotential difference between two remote stations based on clock-transportation experiment. In this work, the experiment yielded large observed data sets, with the valid zero-baseline measurement lasting for 6~days, and the geopotential difference measurement lasting for 65~days,  which is important for verifiing the reliability of the results. In addition, we use the ensemble empirical mode decomposition (EEMD) technique to effectively determine the geopotential-related signals from the CVSTT observations. To our knowledge, this is the first application to CVSTT data processing for determining a geopotential difference. In this study, the stabilities of hydrogen clocks used are at the $10^{-15}$~day$^{-1}$ level, which means that the determined geopotential difference is limited to tens of meters in equivalent height.

\section{Experiment and data processing}\label{sec 2}
\subsection*{Experimental setup}\label{sec 2.1}
\begin{figure*}[h!]
	\centering
	\includegraphics[width=1\textwidth]{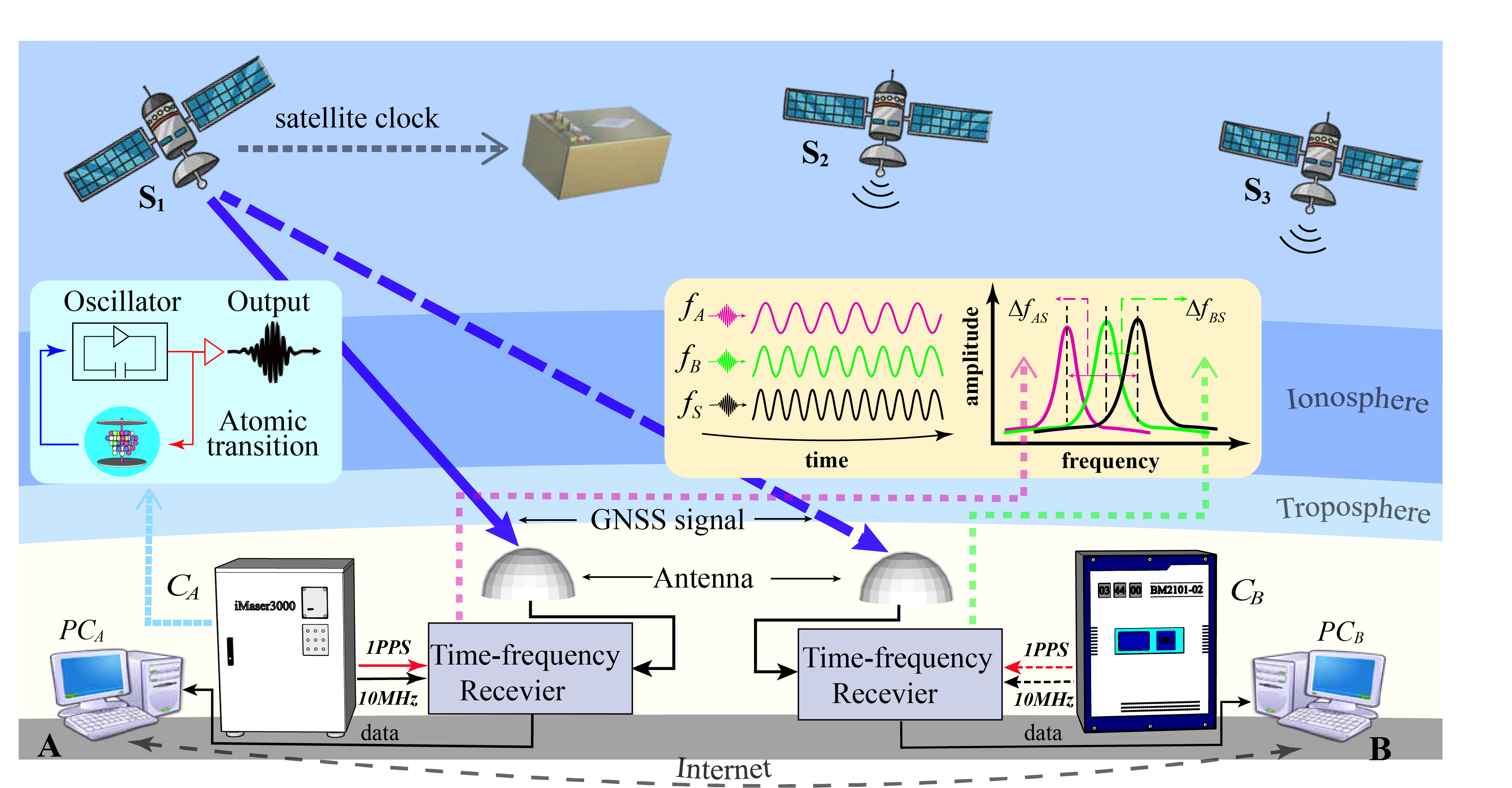}
	\caption{Frequency comparison between two atomic clocks, $C_{_A}$and  $C_{_B}$ at sites A and B, respectively. The atomic clock at the $i$-th station ($i=$A or B) outputs 10~MHz microwave signals and 1 pulse per second (1~PPS) signals, originated from high-stability oscillator with inherent frequency $f_{i}$, and these signals are used for comparing with the corresponding signals from satellite. The GNSS signals containing the frequency $f_{_S}$ of the satellite's clock, are received by the antennas, and then are transfered to the corresponding GNSS time-frequency receivers. The frequencies of the clocks at the $i$-th station and the satellite $S$, $f_{_i} (i=A, B, S)$, are compared at the corresponding receivers, and the results (which contain the information of frequency difference, $\Delta f_{_{AS}}=f_{_S}-f_{_A}$ and $\Delta f_{_{BS}}=f_{_S}-f_{_B}$, between satellte and ground stations ) are transferred to computers $PC_A$ and $PC_B$ via cables, respectively. By taking the common-view satellite, $S_k$ ($k$=1,2,3,$\dots$), as common reference, the clock comparisons between the two ground stations, A and B, can be finally determined. Here, the content inside the light cyan rectangle is the simplified principle of atomic clock, which explains that the high stability frequency generated by the  atomic transition is output after being adjusted by the oscillator; and the content inside  the light yellow rectangle is the simplified principle of clock comparison. More detail about the CVSTT technique are available in SI Appendix.}
\label{fig:1}
\end{figure*}

In this study, the frequency comparison of two hydrogen atomic clocks, one fixed reference clock, $C_{A}$ (iMaser3000), and one portable clock, $C_{B}$ (BM2101-02), are conducted via the CVSTT technique, which is shown in  Fig.~\ref{fig:1}, and the error sources  are given in Table~\ref{tab:1}. Here we use the modified Allan deviation (MDEV) to evaluate the frequency stabilities of the clocks \cite{Riley.2008}, and the coresponding results are given in Table~\ref{tab:2}.
\begin{table}
\centering
\setlength{\tabcolsep}{4mm}
\caption{The error sources of CVSTT technique in the clock-transportation experiment with duration of 6 days.}\label{tab:1}
\begin{tabular}{lccc}
\toprule
Error type & Sources errors & CVSTT accuracy (ns)
\\
\midrule
Broadcast ephemeris         & 2~m & 0.16 &
\\
Ground station              &3~cm & 0.15 &
\\
Ionosphere                  & - & 0.80 &
\\
Troposphere                 & - & 0.52 &
\\
Sagnac                      & $1.4\times10^{-18}$   & 0.003
\\
Total                       &  -   &1.64 &
\\
\bottomrule
\end{tabular}
\end{table}

\begin{table}[!h]
\footnotesize
\centering
\setlength{\tabcolsep}{0.5mm}
\caption{Nomina stabilities of the two hydrogen atomic clocks in the experiiment. $C_{A}$ is the  fixed clock (iMaser3000), $C_{B}$ is the transportable clock (BM2101-02).}\label{tab:2}
\begin{tabular}{cccccc}
\toprule
\multicolumn{1}{c}{\multirow{2}{1cm}{Clock}}
&\multicolumn{5}{c}{Time interval}
\\
\cline{2-6}
\multicolumn{1}{c}{}
&\multicolumn{1}{c}{1~s}&{10~s}&{100~s}&{1000~s}&{10000~s}
\\
\midrule
$C_{A}$ & $1.50 \times 10^{^{-13}}$ & $2.00 \times 10^{^{-14}}$ & $5.00 \times 10^{^{-15}}$ & $2.00 \times 10^{^{-15}}$ & $2.00 \times 10^{^{-15}}$
\\
$C_{B}$ & $4.57 \times 10^{^{-13}}$ & $8.85 \times 10^{^{-14}}$ & $1.95 \times 10^{^{-14}}$ & $5.96 \times 10^{^{-15}}$ & $2.18 \times 10^{^{-15}}$
\\
\bottomrule
\end{tabular}
\end{table}
\begin{figure*}[h!]
	\centering	\includegraphics[width=1\textwidth]{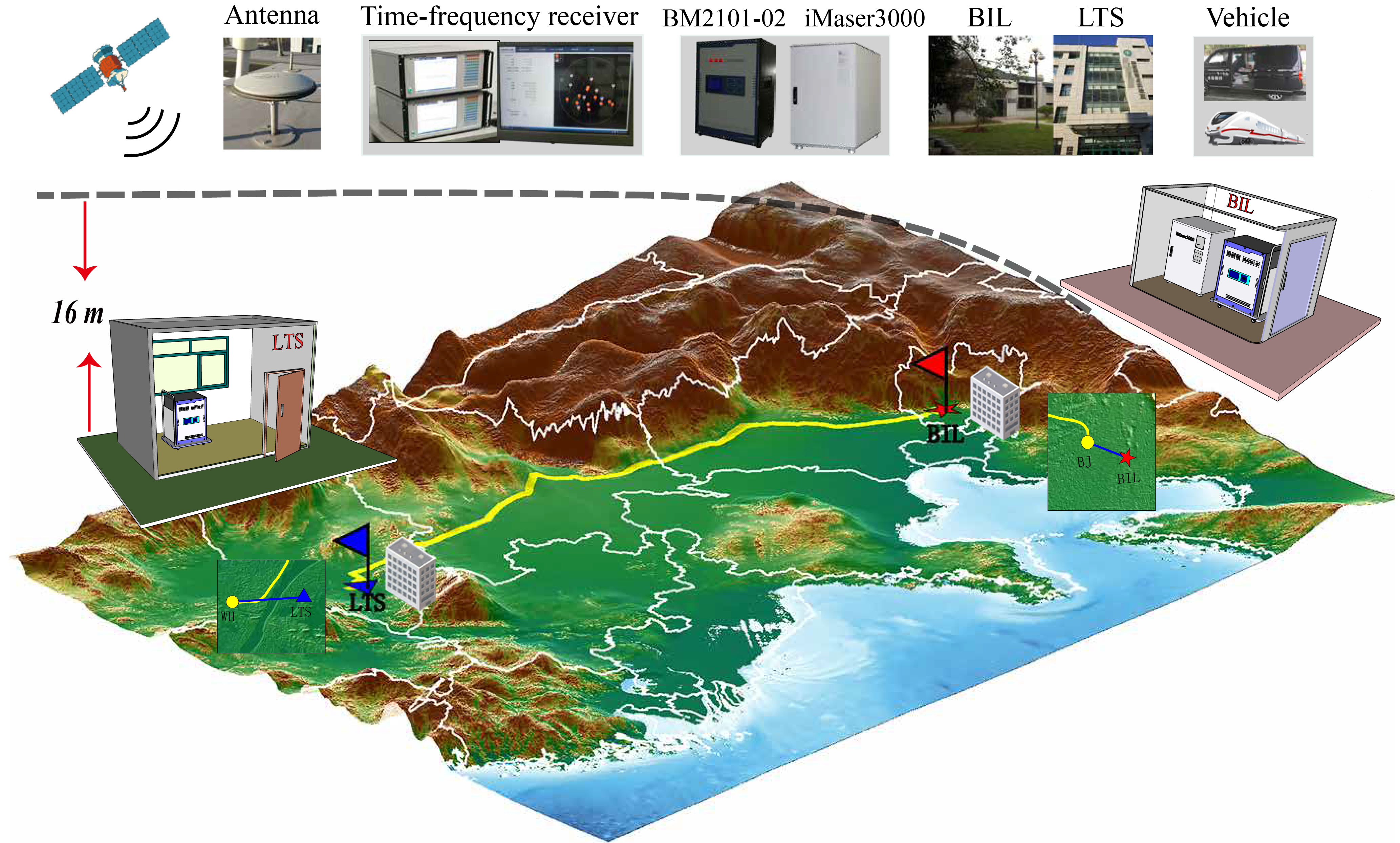}
	\caption{Schematic diagram of the clock-transportation experiment. In Period~1, the zero-baseline measurement was implemented with both $C_{A}$ and $C_{B}$ located at BIL, from January 13 to 29, 2018. In Period~2, after clock $C_{B}$ was transported to LTS from BIL, the geopotential difference measurement was completed by $C_{A}$ and $C_{B}$ located at BIL and LTS, from February 01 to April 07, 2018. The BIL and LTS are around 1000~km apart, and the orthometric heihgt (OH) difference of the clocks between BIL and LTS is about 16~m. The clock $C_{B}$ was transported via high-speed train (route denoted as yellow) and the dedicated experimtnal vehicle (route denoted as blue).}
	\label{fig:2}
\end{figure*}
\subsection*{Experimental process}\label{sec 2.2}
The experiment was done at the Beijing 203 Institute Laboratory (BIL) and Luojiashan Time--Frequency Station (LTS), with a distance around 1000~km apart, and the specific information are shown in Fig.~\ref{fig:2} and Table~\ref{tab:3}. The experiment is divided into two periods, during which the zero-baseline measurements (denoted as Period~1) and the geopotential difference measurements (denoted as Period~2) were obtained. The purpose of Period~1 is to estimate the constant systematic shift of the two-clock system, while Period~2 is used to measure the frequency difference caused by the geopotentials. In Period~1 spanning from  January 13 to 29, 2018, the zero-baseline measurements were made at BIL with clocks $C_{A}$ and $C_{B}$ being placed in the same shielding room under the same geopotential. According to the GNSS tracking schedule, the clock comparisons between $C_{A}$ and $C_{B}$ were conducted with the CVSTT technique, and a series of time difference, $\Delta t_{_{AB}}(i)$, were obtained, where $\Delta t_{_{AB}}(i)=t_{_{B}}(i) -t_{_{A}}(i)$.

\begin{table}
\footnotesize
\centering
\setlength{\tabcolsep}{10mm}
\caption{Detailed information of the two ground stations (BIL and LTS) in the clock transportation experiment. The coordinates ($\varphi$, $\lambda$, $h$) denote the locations of the GNSS antennas (under the frame of WGS 84). $H_{clock}$ denote the OHs of the hydrogen clock in two staions, respectively. $H_{clock}$ is determined by the EGM2008 gravity field model and a tape measure.}\label{tab:3}
\begin{tabular}{ccccc}
\toprule
\multicolumn{1}{c}{\multirow{2}{0.8cm}{Stations}}
&\multicolumn{3}{c}{Antennas location}
&\multicolumn{1}{c}{\multirow{2}{0.8cm}{$H_{clock}$~(m)}}
\\
\cline{2-4}
\multicolumn{1}{c}{}
&\multicolumn{1}{c}{$\varphi$~($\circ$)}&$\lambda$~($\circ$)&{$h$~(m)}
&\multicolumn{1}{c}{}
\\
\midrule
BIL &116.26 & 39.91 & $70.51$ & $51.55$ 
\\
LTS &114.36 & 30.53 & $28.00$ & $35.51$ 
\\
\bottomrule
\end{tabular}
\end{table}
\begin{figure*}[h!]
	\centering
	\includegraphics[width=1\textwidth]{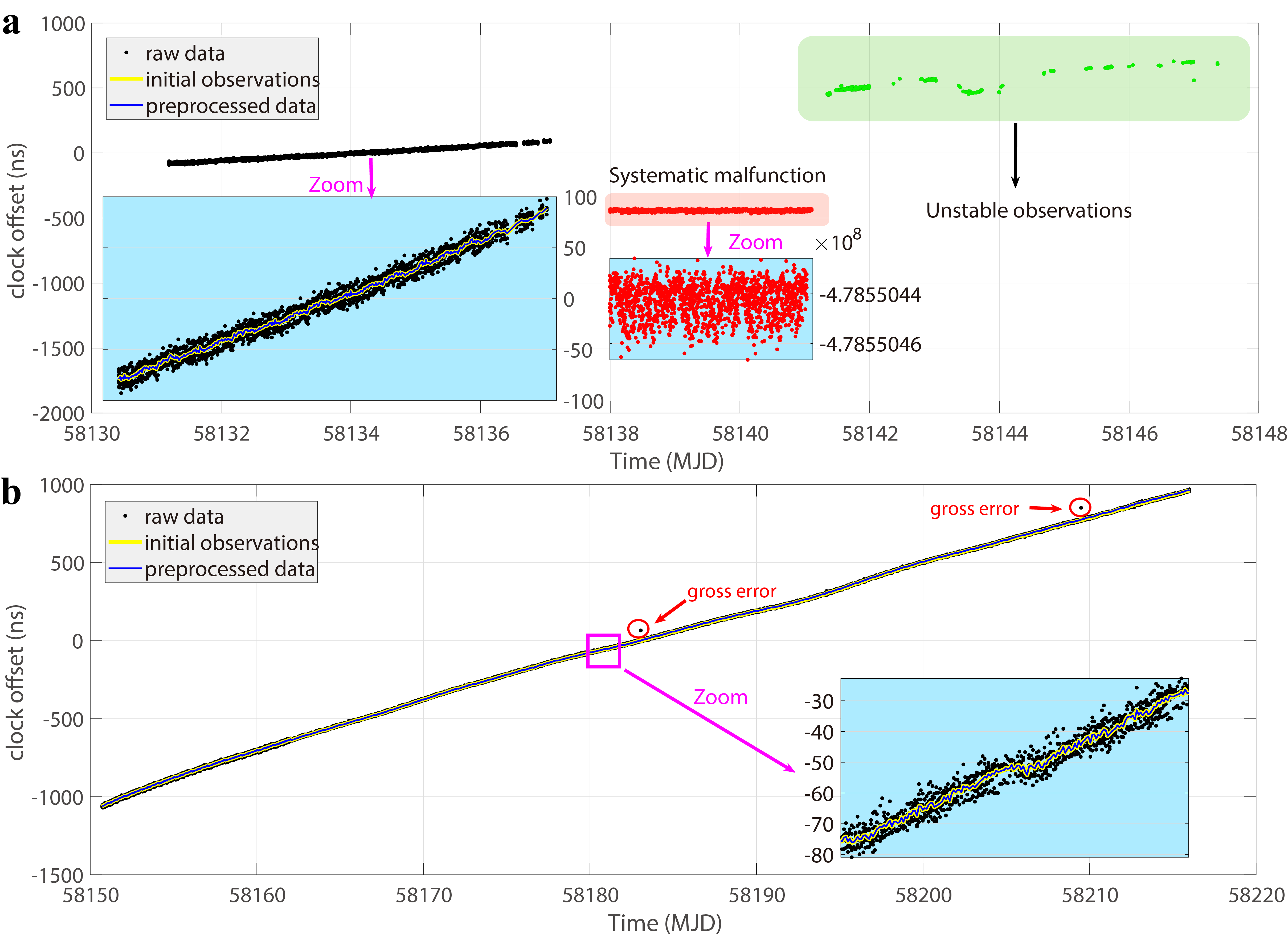}
	\caption{The time difference series between two hydrogen clocks $C_{A}$ and $C_{B}$, $ \Delta t_{_{AB}}(t)$, obtained with  CVSTT technique in (a) Period~1 and (b) Period~2. The raw data (black points) denote the discrete observations from different common-view satellites. The initial observations (yellow curve) denote the initial time difference series, which are the mean of raw data. The preprocessed data (blue curve) denote the preprocess time difference series after gross error removal, clock jump correction and missing data interpolation  on the initial observations. In Period~1 (a), there is a systematic malfunction from MJD 58138 to 58141 (red points in light red rectangle, the data have increased by 478,550,000~ns for the effective visualization), and the observations are unstable with lots of missing data from MJD 58141 to 58147 (green points in light green rectangle).}
	\label{fig:3}
\end{figure*}
\begin{figure*}[h!]
	\centering
	\includegraphics[width=1\textwidth]{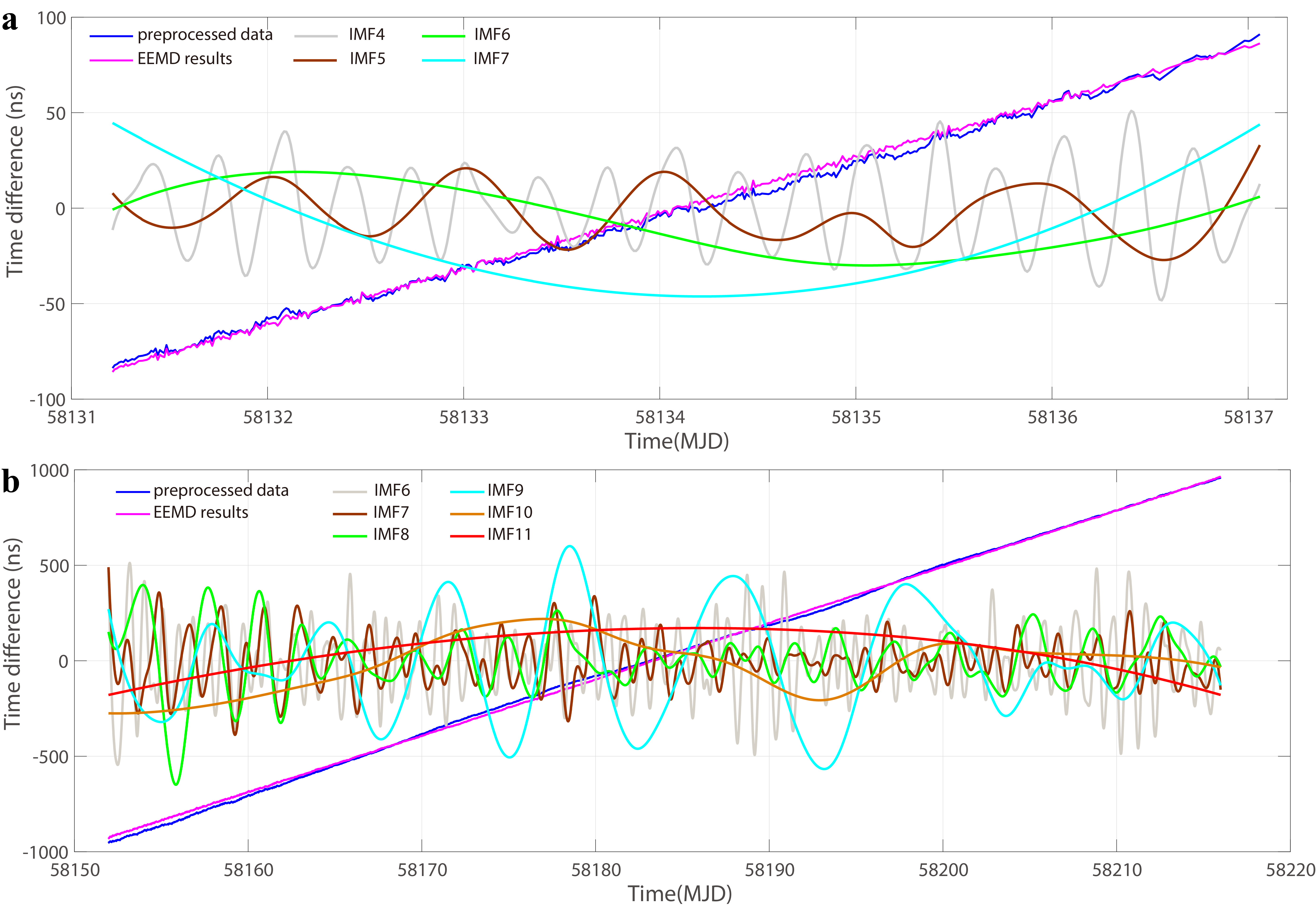}
	\caption{The time difference series before and after EEMD technique in (a) Period~1 and (b) Period~2. The "preprocessed data" (blue curve) is the preprocessed data set, $\Delta t_{_{AB}}(t)$, and the "after-EEMD data" (purple curve) is the reconstructed data set from $\Delta t_{_{AB}}(t)$ based on EEMD technique, denoted as $\Delta t_{_{AB}}^{re}(t)$. The IMFs denoted in different colors are the corresponding removed periodic components in Period~1 and Period~2, respectively. It is noted here that to aid visualization, the removed IMFs in (a) have been amplified by a factor of 20; and the removed IMFs in (b) have been amplified by 200, except for IMF10, which is amplified by 10.}
	\label{fig:4}
\end{figure*}

When Period~1 was finished, the portable clock, $C_{B}$, was transported to LTS from BIL while clock $C_{A}$ remained at BIL. During transportation the portable clock, $C_{B}$, continued operating. After installation and adjustment of $C_{B}$ at LTS, clock comparisons between $C_{A}$ and $C_{B}$ were conducted during Period~2, which spanned from February 01 to April 07, 2018. It should be noted that there was no difference in the experimental setup throughout the entire experiment except for the different location of clock $C_{B}$ during Period~1 and Period~2. In addition, there are about two days of missing data (from January 30 to 31, 2018) before re-observation due to the clock transportation and experimental installation at station LTS.

As temperature is a major factor which may disturb the performance of atomic clocks, we controlled the environment with a relatively constant temperature throughout the entire experiment. The temperature was held nearly constant ($24\pm0.5^{\circ}C$) for the fixed clock, $C_{A}$, in Period~1 and Period~2 at BIL. For portable clock, $C_{B}$, it was in the same laboratory with clock $C_{A}$ during Period~1, and the temperature was ($24\pm1.5^{\circ}C$) in Period~2 at LTS.

\subsection*{Data Processing}\label{sec: 2.3}
Based on the observations via the CVSTT technique, the time difference  series, $\Delta t_{_{AB}}(t)$, in Period~1 and Period~2  are determined, respectively, referred to Fig.~\ref{fig:3} , in which, the time difference  series, $\Delta t_{_{AB}}(t)$, of the observations for different common-view satellites are denoted as "raw data" (black points). Next, the initial CVSTT observations are obtained based on the "raw data", which are denoted as "initial observations" (yellow curve). After that, we performe data preprocess on the initial observations by removing gross errors, correcting clock jumps, and inserting missing data, and the preprocessed CVSTT data are finally obtained, which are denoted as "preprocessed data" (blue curve). In the following data processing, all analysis are based on the preprocessed data sets. Here, it should be noted that the observations from MJD 58138 to 58141 are not used due to a systematic malfunction, and the observations from MJD 58141 to 58147 are also not used due to unstable observations (too many missing observations). Therefore, the valid observations in Period~1 span from MJD 58131 to 58136 (as seen in Fig.~\ref{fig:3}(a)).

Next, the EEMD technique \cite{Wu.2009b} is applied to process (filter) the  preprocessed data sets for effectively determine the geopotential-related signals. The reason is as follows.
Our main objective in this study is to measure the frequency shift between two clocks caused by geopotential difference, which has linear characteristic in time difference series in the experiment, due to the fact that the stations of BIL and LTS are fixed. However, the CVSTT observations inevitably contain periodic signals (e.g. influences by temperature), which are the interfering signals for our target signals. For this reason, the periodic components included in  $\Delta t_{_{AB}}(t)$ should be removed. 

By using EEMD technique, the preprocessed data sets, $\Delta t_{_{AB}}(t)$, in Period~1 and Period~2 are decomposed into a series of intrinsic mode functions (IMFs, with frequencies from higher to lower) and a long trend component, $r$, respectively (SI Appendix; Fig.~S6 and S7). Then, we examine the completeness and orthogonality of the EEMD decomposition, with the index of orthogonality (IO). If the decomposition is completely orthogonal, which indicates that IO= 0, and for the worst case, IO= 1. The IO values corresponding to Period~1 and Period~2 are $0.0085$ and $0.0094$, respectively, demonstrating that the signals are effectively decomposed. Finally, the uninteresting periodic components included in the $\Delta t_{_{AB}}(t)$ are removed (SI Appendix, Fig.~S8), and we reconstruct the time difference series, $\Delta t_{_{AB}}^{re}(t)$, by summing the residual components, which is denoted as "after-EEMD data". The time difference series before and after application of the EEMD technique, $\Delta t_{_{AB}}(t)$  and  $\Delta t_{_{AB}}^{re}(t)$, are shown in Fig.~\ref{fig:4}, and the corresponding MDEVs are shown in SI Appendix (see Fig.~S10). The $\Delta t_{_{AB}}^{re}(t)$ are considered as containing only geopotential-related signals, based on which we determine the geopotential difference between $C_{B}$ (LTS) and $C_{A}$ (BIL).

Concerning the zero-baseline measurement, i.e., Period~1, after EEMD decomposition and removing periodic IMFs, the reconstructed time difference series, $\Delta t_{_{AB}}^{re}(t)$, and the corresponding frequency shift $\Delta f_{zero}/f_{0}$ is then determined. We take this result as a constant systematic shift in the experiment. For the geopotential difference measurement, i.e., Period~2, we implement a grouping strategy by segmenting the corresponding $\Delta t_{_{AB}}^{re}(t)$ into successive six-day measurements as independent segments with no data overlap, for matching with the six-day zero-baseline duration as well as for improving the data utilization. Furthermore, with this strategy, the clock drift can be limited to a relative short time duration (i.e., six days). The reconstructed time difference series, $\Delta t_{_{AB}}^{re}(t)$, and corresponding MDEVs of each segment in Period~2 can be seen in SI Appendix (Figs.~S9 and S10). Based on each segment, a series of frequency shift $\Delta f_{geo}(i)/f_{0}$ ($i$-th segment) is determined. Then, by taking the results of $\Delta f_{zero}/f_{0}$ in Period~1 as constant systematic shift, the net frequency shift caused by the geopotential difference between stations LTS and BIL is finally determined as $\Delta f/f_{0}=<\Delta f_{geo}(i)/f_{0}-\Delta f_{zero}/f_{0}>$, where $"<>"$ denote the weighted average results based on corresponding MDEVs, which means that for a certain segment, the higher the observation stability, the greater the weight. The experiment results of the net frequency shift determined in each segment, $\Delta f(i)/f_{0}$, are shown in Fig.~\ref{fig:5}.

\section*{Results}\label{sec:3}
\begin{figure*}[h!]
\centering
\includegraphics[width=1\textwidth]{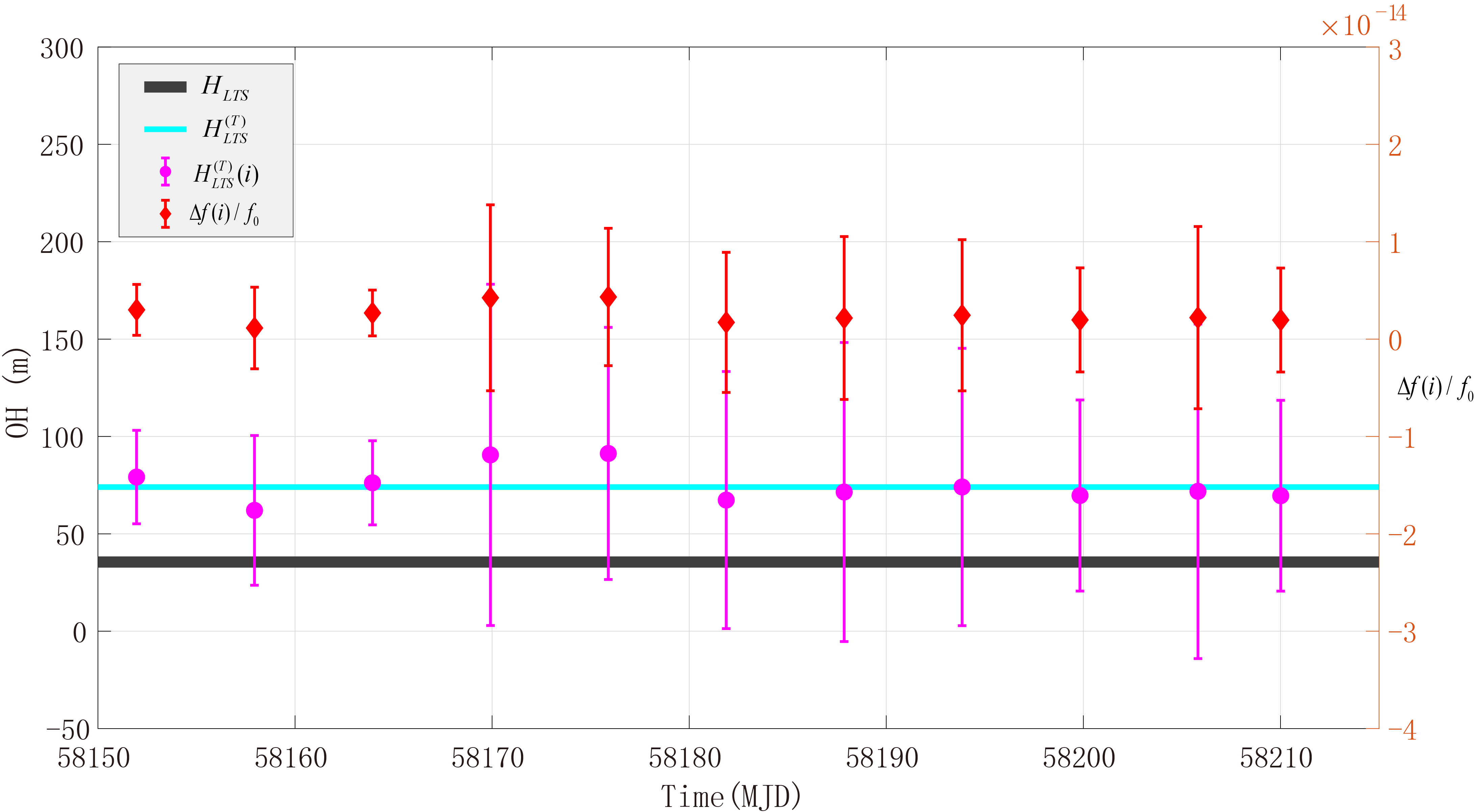}
\caption{The experiment results of frequency shift and OH determination of LTS station. $\Delta f(i)/f_0$ (red scatter) denote the net frequency shift between stations LTS and BIL determined in each segment, respectively, with $\Delta f(i)/f_0=\Delta f_{geo}(i)/f_0-\Delta f_{zero}/f_0$, $H_{LTS}^{_{(T)}}(i)$ (purple scatter) denote the corresponding clock-comparison-determined OH at LTS, $H_{LTS}^{_{(T)}}$ (cyan curve) denote the weighted mean results of $H_{LTS}^{_{(T)}}(i)$. $H_{LTS}$ (grey curve) denote the OH of station LTS determined by EGM2008, which is as a comparison.}
\label{fig:5}
\end{figure*}

According to Eq. \eqref{eq:0}, the clock-comparison-determined geopotential difference, $\Delta W_{_{AB}}$, as well as the OH of station LTS, $H_{_{LTS}}^{_{(T)}}$, are finally determined, respectively, as shown in Fig.~\ref{fig:5} and Table~\ref{tab:4}. The corresponding specific results are ($220\pm448$)~m$^{2}$s$^{-2}$ and ($74.0\pm45.7)$~m, respectively. Compared to the corresponding EGM2008 model result, which suggests that $H_{LTS}=35.5$~m, the deviation between the clock-comparison-determined result and the model value is 38.5~m. The results are consistent with hydrogen atomic clocks used in our experiment with the frequency stabilities at $10^{-15}$ day$^{-1}$ level.

\begin{table*}
\footnotesize
\centering
\setlength{\tabcolsep}{0.2mm}
\caption{The experiment results of frequency shift between stations LTS and BIL, $\Delta f(i)/f_0$, and corresponding OH determination of LTS station, $H_{LTS}^{_{(T)}}(i)$ (unit: m). "<>" denote the weighted average result.   
Unit of $\Delta f_{zero}/f_0$, $\Delta f_{geo}(i)/f_0$ and $\Delta f(i)/f_0$: $10^{-15}$.}\label{tab:4}
\begin{tabular}{ccccccccccccc}
\toprule
\multicolumn{6}{c}{$\Delta f_{zero}/f_0=338.3\pm1.0$}
&\multicolumn{6}{c}{$H_{LTS}=35.51$~m}
\\
\midrule
\multicolumn{1}{r}{\multirow{2}{1.5cm}{Experiment results}}
&\multicolumn{11}{c}{Segment in Period~2}
&\multicolumn{1}{c}{\multirow{2}{0.2cm}{<>}}
\\
\cline{2-12}
\multicolumn{1}{c}{}
&\multicolumn{1}{c}{1th}& {2th}& {3th}& {4th}& {5th}& {6th}& {7th}& {8th}& {9th}& {10th}& {11th}
&\multicolumn{1}{c}{}
\\
\midrule
$\Delta f_{geo}(i)/f_0$  &341.3$\pm$2.4  & 339.4$\pm$4.1 & 341.0$\pm$2.1 & 342.5$\pm$9.5 & 342.6$\pm$7.0 & 340.0$\pm$7.1 & 340.1$\pm$8.3 & 340.7$\pm$7.7 & 340.3$\pm$5.3 & 340.5$\pm$9.3 & 340.3$\pm$5.3 &-
\\
$\Delta f(i)/f_0$  & 3.0$\pm$2.6  & 1.1$\pm$4.2 & 2.7$\pm$2.4 & 4.2$\pm$9.6 & 4.3$\pm$7.1 & 1.7$\pm$7.2 & 2.2$\pm$8.4 & 2.5$\pm$7.8 & 2.0$\pm$5.3 & 2.2$\pm$9.4 & 2.0$\pm$5.3 &2.4$\pm$5.0
\\
$H_{_{LTS}}^{_{(T)}}(i)$ & 79.2$\pm$24.0 & 62.1$\pm$38.5 & 76.2$\pm$21.6 & 90.6$\pm$87.6 & 91.3$\pm$64.7 & 67.3$\pm$66.0 & 71.5$\pm$76.8 & 74.1$\pm$71.2 & 69.7$\pm$49.1 & 71.8$\pm$85.9 & 69.6$\pm$49.0 &74.0$\pm$45.7
\\
\bottomrule
\end{tabular}
\end{table*}
\section*{Discussion} \label{sec:4}
In this study, an experiment of clock comparison based on the CVSTT technique for determining the geopotential difference is investigated. Based on this approach, one can directly determine the geopotential difference between two ground points. The experimental results shows that the clock-comparison-determined OH result, $H_{_{LTS}}^{_{(T)}}$,  is ($74.0\pm 45.7$)~m. The discrepancy between the experiment result and the EGM2008 model value is 38.5~m. This is consistent with the frequency stabilities of the hydrogen atomic clocks used in the experiment.

Here, we first use the EEMD technique on the CVSTT observations to determine the geopotential-related signals more effectively. In addition, we use about 71 days of observations to determine the geopotential difference between two stations based on the CVSTT technique,   and sufficient observations have good advantage in verifiing the reliability of the approach, which could be applied extensively in geodesy in the future.

In this study, the deformation of the Earth's surface caused by tides is neglected because the tidal influences are at the tens of centimeters level, comparing with the stabilities of the hydrogen atomic clocks used in the experiment being at the $10^{-15}$~day$^{-1}$ level, which is equivalent to tens of meters in OH. The residual errors caused by the ionosphere, troposphere, and Sagnac effects were also neglected because these errors are far below the accuracy level of the hydrogen clocks (see Table~\ref{tab:1}). However, to achieve centimeter-level measurement in the future, the above mentioned influences should be taken into consideration carefully. In addition, for the purpose of centimeter-level measurements, we need to use the gravity frequency shift equation accurate to the $c^{-4}$ level \cite{Shen.2017}.

With the rapid development of time and frequency science and technology, the approach discussed in this study is promising, and could be implemented as an alternative technique for establishing height datum networks. In addition, it could be applied to unifying the WHS. Using better atomic clocks with higher stabilities may significantly improve the results with higher accuracy.\\

\section*{Methods}
\subsection*{Principle}
We determine the geopotential by comparing the frequencies of two remote clocks. The principle is stated as follow. Considering two identical  clocks $C_{A}$ and $C_{B}$ located at two ground stations A and B with the geopotentials $W_{A}$ and $W_{B}$, the observed frequencies are $f_{_A}$ and $f_{_B}$, respectively. Based on the gravity frequency shift equation  $\Delta f_{_{AB}}/f_{_0}=-\Delta W_{AB}/c^{2}$  \cite{Bjerhammar.1985, Mehlstaubler.2018,  Cai.2020}, we may determine the geopotential difference $\Delta W_{AB}=W_{B}-W_{A}$, where $\Delta f_{_{AB}}=f_{_B}-f_{_A}$, is the frequency difference between $C_{A}$ and $C_{B}$, $c$ is the speed of light in vacuum, $f_{0}$ is inherent frequency of the clocks. Here, we used the conventional definition of the geopotential in geodesy ($W_i$ is possitive).

Suppose the geopotential $W_{A}$ at station A is a priori given, accurate to $c^{-2}$, which is equivalent to several centimeters on the ground, the clock-comparison-determined OH at station B, $H_{_B}^{_{(T)}}$, can be determined as \cite{HofmannWellenhof.2006, Jekeli.2000, Wu.2021}: 
\begin{center}
	\begin{equation}\label{eq:0}
	H_{_B}^{_{(T)}}=\frac{W_{0}-W_{A}}{\bar g_{_B}}+\frac{c^{2}}{\bar g_{_B}}(\frac{\Delta f_{_{AB}}}{f_{_0}})
	\end{equation}
\end{center}
where $W_{0}$ is the geopotential on the geoid, with $W_{0}=$ 62,636,853.35 $\pm$0.02~m$^{2}$~s$^{-2}$ \cite{Sanchez.2016}, $\bar{g}_{_B}$ is the mean gravity value of station B.

\subsection*{CVSTT technique}
We used the CVSTT technique for comparing the frequencies between two hydrogen clocks. The main advantages of the CVSTT technique lie in that the satellite clock errors are cancelled, and various other errors, especially ionospheric and tropospheric errors, are largely reduced due to the simultaneous two-way observations. The principle as well as various kinds of errors in CVSTT technique are described in SI Appendix.

\subsection*{EEMD technique}
We used the EEMD technique \cite{Wu.2009b} to the CVSTT observations for more effectively determining the geopotential-related signals from the time difference series of the two clocks. Details about the EEMD technique are presented and discussed in SI Appendix.

\section*{Acknowledgement}
This study was supported by the National Natural Science Foundations of China (Nos. 42030105, 41721003, 41804012, 41631072, and 41874023), Space Station Project (2020) (No. 228), and the Natural Science Foundation of Hubei Province of China (No. 2019CFB611). We thank International Science Editing (http://www.internationalscienceediting.com) for editing this manuscript.


\newpage


\makeatletter 
\renewcommand{\thefigure}{S\@arabic\c@figure}
\renewcommand{\theequation}{S\@arabic\c@equation}
\makeatother
\setcounter{figure}{0}
\setcounter{equation}{0}

\section*{Supplementary information Appendix}
The supplementary information (SI) aims at illustrating the principle and effectiveness of the common view satellite time transfer (CVSTT) technique and ensemble empirical mode decomposition (EEMD) technique used in this study, and give some results of the experimental results.
\section*{Overview of the CVSTT Technique}\label{sec 1}
\subsection*{CVSTT technique}\label{sec 1.1}
Here we briefly introduce the principle of the CVSTT technique. Suppose there are two ground stations, A and B, and a common-view satellite, S. The coordinates of the two ground stations are denoted as $(x_{i},y_{i},z_{i})$ ($i$=A or B), and the satellite coordinates are denoted as ($x_{_S}$, $y_{_S}$, $z_{_S}$). The simultaneous observations for satellite S are conducted at stations A and B through common-view receivers, with the help of the global navigation satellite system (GNSS). Therefore, for the $i$-th station and an appointed time, $\tau$, the observation model can be constructed as follows \cite{Defraigne.2011, ElliottD.Kaplan.2017, Ansari.2015}: 
\begin{equation}\label{eq:1}
\rho_{si}=r_{si}+I_{i}+T_{i}+Sag_{i}+c(\delta t_{i}-\delta t_{_S})+t_{r_{i}}+\varepsilon_{i}
\end{equation}
where $\rho_{si}$ is the precise code measurement of the $i$-th station, $I_{i}$, $T_{i}$, and $Sag_{i}$ are the ionospheric delay error, tropospheric delay error, and Sagnac effect error of the $i$-th station, respectively, $t_{r_{i}}$ is the receiver delay, which is usually calibrated by the manufacturing company using simulated signals, $\varepsilon_{i}$ is the residual noise error, and $r_{si}=\sqrt{(x_{i}-x_{_S})^{2}+(y_{i}-y_{_S})^{2}+(z_{i}-z_{_S})^{2}}$ is the geometric distance between satellite $S$ and the $i$-th station, $\delta t_{i}$ ($\delta t_{_S}$) is the time difference between the $i$-th station (satellite S) and the common reference time, i.e., the GPS time ($t_{_{GPS}}$), expressed as \cite{ElliottD.Kaplan.2017}:
\begin{equation}\label{eq:2}
\begin{split}
\delta t_{i}=t_{i}-t_{_{GPS}}\\
\delta t_{_{S}}=t_{_{S}}-t_{_{GPS}}
\end{split}
\end{equation}
where $t_{i}$ and $t_{_{S}}$ are the clock records at reference time $t_{_{GPS}}$.

By combining Equations~(\ref{eq:1}) and (\ref{eq:2}), the time difference between the $i$-th station and satellite S at time $\tau$ can be determined as:
\begin{equation}\label{eq:3}
\begin{split}
c\Delta t_{_{SA}}=\rho_{_{SA}}-r_{_{SA}}-I_{_A}-T_{_A}-Sag_{_A}-t_{r_{_{A}}}-\varepsilon_{_A}\\
c\Delta t_{_{SB}}=\rho_{_{SB}}-r_{_{SB}}-I_{_B}-T_{_B}-Sag_{_B}-t_{r_{_{B}}}-\varepsilon_{_B}
\end{split}
\end{equation}

Therefore, at the appointed time, $\tau$, the time difference between the two ground stations, A and B, can be determined by taking the common-view satellite as a common reference:
\begin{equation}\label{eq:4}
\begin{split}
\Delta t_{_{AB}}&=t_{_{B}}-t_{_{A}}\\
&=
\frac{\rho_{_{SB}}-\rho_{_{SA}}}{c}-\frac{r_{_{_{SB}}}-r_{_{_{SA}}}}{c}-\frac{I_{_B}-I_{_A}}{c}-\frac{T_{_B}-T_{_A}}{c}\\
&-\frac{Sag_{_B}-Sag_{_A}}{c}-\frac{t_{r_{_{B}}}-t_{r_{_{A}}}}{c}-\frac{\varepsilon_{_B}-\varepsilon_{_A}}{c}\\
\end{split}
\end{equation}

 We note that, the observed $\rho$ is derived from the code phase\cite{ElliottD.Kaplan.2017}. Hence, the CVSTT technique is essentially a phase comparison or frequency comparison between remote clocks, because the derivative of the phase with respect to time is frequency. And by continuous comparison, the time difference at each time $t (\tau_{1},\tau_{2},\tau_{3},\cdots)$ can be determined, which construct a time difference series, $\Delta t_{_{AB}}(t)$. In this study, we conduct an experiment of determining the gravity potential with the CVSTT technique using two portable hydrogen clocks. The experiment was conducted at the Beijing 203 Institute Laboratory (BIL) and Luojiashan Time--Frequency Station (LTS). In the sequel, we will analyze various error sources.

\subsubsection*{Satellite position errors}\label{sec 1.2.1} 
The accuracy of a satellite position depends on its ephemeris. Suppose the satellite position error is ($\delta x_{s}$, $ \delta y_{s}$, $ \delta z_{s}$). Ignoring the position error of ground stations, the relation between the satellite position error and time-transfer error in the CVSTT technique can be determined by using the first-order difference \cite{Imae.2004, Sun.2010}:
\begin{equation}\label{eq:5}
\Delta t^{_{SP}}_{_{AB}}=(l_{_{AS}}-l_{_{BS}})\frac{\delta x_{s}}{c}+(m_{_{AS}}-m_{_{BS}})\frac{\delta y_{s}}{c}+(n_{_{AS}}-n_{_{BS}})\frac{
	\delta z_{s}}{c}
\end{equation}
where
\begin{equation}\label{eq:6}
l_{is}=\frac{x_{s}-x_{i}}{\rho_{is}},\hspace{2mm}
m_{is}=\frac{y_{s}-y_{i}}{\rho_{is}},\hspace{2mm} n_{is}=\frac{z_{s}-z_{i}}{\rho_{is}}.  
\end{equation}

Based on Equations~(\ref{eq:5}) and (\ref{eq:6}), and applying the error-propagation law, the following expression can be obtained: 
\begin{equation}\label{eq:7}
\begin{split}
m^{2}(\Delta t^{_{SP}}_{_{AB}})&=(\frac{l_{_{AS}}-l_{_{BS}}}{c})^{2}m^{2}(\delta x_{s}) +(\frac{m_{_{AS}}-m_{_{BS}}}{c})^{2}m^{2}(\delta y_{s})\\
&+
(\frac{n_{_{AS}}-n_{_{BS}}}{c})^{2}m^{2}(\delta z_{s})\\
\end{split}
\end{equation}

For instance, when the accuracy of the  broadcasted ephemeris for an individual axis is 2~m, the error of the time transfer caused by the satellite positional error did not exceed 0.16~ns in the experiments. The influences of the ground stations' position errors on the time transfer are evaluated similarly. Here the ground stations' coordinates were determined based on the precise point positioning method in advance, with an accuracy level of better than 3~cm \cite{Dawidowicz.2014}, and the corresponding time-transfer error did not exceed 0.15~ns.

\subsubsection*{Ionospheric effects}\label{sec 1.2.2}
The ionospheric effect on a GNSS signal transmision could be up to several tens of nanoseconds, due to the electron content of the atmospheric layer; the effect becomes more dramatic during an ionospheric storm. The zenith group delays are typically about 1$\sim$30~m, 0$\sim$2~cm, and 0$\sim$2~mm for the first-, second-, and third-order ionospheric effects, respectively \cite{Marques.2011b}. For the GNSS time-transfer receivers, measurements on two frequencies ($f_{1}$ and $f_{2}$) are often available. Therefore, an ionosphere-free observation, $P_{3}$, can be constructed so that the first-order term of the ionospheric effect can be removed completely, due to the fact that the ionospheric effect on a signal depends on its frequency. $P_{3}$ is constructed as follows \cite{Petit.2009}:
\begin{equation}\label{eq:8}
P_{3}=\frac{1}{f_{1}^2-f_{2}^2}(f_{1}^2P_{1}-f_{2}^2P_{2})
\end{equation}
where $P_{1}$ and $P_{2}$ are the precise-code observations with frequencies of $f_{1}=1575.42$~MHz and $f_{2}=1227.60$~MHz, respectively. 

However, Eq. \eqref{eq:8} only removes the first-order ionospheric effect. There are second- and third-order ionospheric effects (also denoted as higher-order ionospheric effects). The residual higher-order ionospheric effects, $I_{r}$, can be expressed as \cite{Morton.2009}:
\begin{equation}\label{eq:9}
\begin{split}
&I_{r}=\frac{q}{f_{1}f_{2}(f_{1}+f_{2})}+\frac{t}{f_{1}^{2}f_{2}^{2}} \\
&q=2.2566 \times 10^{12}\int N_{e}B_{0}cos\theta_{_B}\,ds \\
&t=2437\int N_{e}^{2}\,ds+ 4.74 \times 10^{22} \int N_{e}B_{0}^{2}(1+cos^{2}\theta_{_B})\,ds \\
\end{split}
\end{equation}
where $N_{e}$ is the number of electrons in a unit volume (m$^{-3}$), $B_{0}$ is the magnitude of the plasma magnetic field (T), $\theta_{_B}$ is the angle between the wave propagation direction and the local magnetic field direction, and $ds$ is the integral element along the path of the wave. With the CVSTT technique, the residual errors of the ionospheric effects between ground stations A and B, $I_{r_{(AB)}}$, can be determined as follows:
\begin{equation}\label{eq:10}
I_{r_{(AB)}}=\frac{q_{_A}-q_{_B}}{f_{1}f_{2}(f_{1}+f_{2})}+\frac{t_{_A}-t_{_B}}{f_{1}^{2}f_{2}^{2}}
\end{equation}

Using the constructed ionosphere-free observation, $P_{3}$, we removed the first-order ionospheric effect, which contributed more than 99\% of all ionospheric effects \cite{Harmegnies.2013}. The residual errors of higher-order terms were about 2$\sim$4~cm (equivalent to 0.07$\sim$0.14~ns)\cite{Morton.2009}. Code noise should also be taken into consideration carefully. Suppose the code noises of $P_{1}$ and $P_{2}$ are $m_{1}$ and $m_{2}$, respectively. By applying the error-propagation law, the code noise of the ionosphere-free observation, $m_{3}$, can be expressed as:
\begin{equation}\label{eq:11}
m_{3}^{2}=(\frac{f_{1}^2}{f_{1}^2-f_{2}^2})^{2}m_{1}^{2}+(\frac{f_{2}^2}{f_{1}^2-f_{2}^2})^{2}m_{2}^{2}
\end{equation}

Combining Equations (\ref{eq:10}) and (\ref{eq:11}), after implementing the strategy of ionosphere-free combination, the residual ionospheric errors, $\Delta t^{ion}_{_{AB}}$, in the CVSTT technique will not exceed 0.8~ns.

\subsubsection*{Tropospheric effects}\label{sec 1.2.3}
The troposphere is the lower layer of the atmosphere that extends from the ground to the base of the ionosphere. The signal transmission delay caused by the troposphere is about 2$\sim$20 m from zenith to the horizontal direction \cite{Chen.2012}. The tropospheric delay depends on the temperature, pressure, humidity, and as well as the location of the GNSS antenna. The total tropospheric delay can be divided into dry and wet parts, which can be expressed as follows \cite{Kouba.2008}:
\begin{equation}
\label{eq:12}
\Delta L=\Delta L_{h}^{z}\cdot mf_{h}(E, a_{h}, b_{h}, c_{h})+\Delta L_{w}^{z}\cdot mf_{w}(E, a_{w}, b_{w}, c_{w})
\end{equation}
where $\Delta L$ is the total tropospheric delay, $\Delta L_{h}^{z}$ and $\Delta L_{w}^{z}$ are the dry and wet parts in the zenith delay, respectively, and $mf_{h}$ and $mf_{w}$ are the corresponding mapping functions (MFs). Each MF is a function of $E$, $a_{h(w)}$, $b_{h(w)}$, and $c_{h(w)}$, where $E$ is the elevation angle in radians, and $a_{h(w)}$, $b_{h(w)}$, and $c_{h(w)}$ are MF coefficients.

A rigorous approach for utilizing numerical weather models (NWMs) for MF determinations was introduced by Boehm and Schuh in 2004\cite{Boehm.2004}. The first and most significant MF coefficients, $a_{h}$ and $a_{w}$, can be fitted with the NWM from the European Centre for Medium-Range Weather Forecasts (ECMWF). Coefficient $b_{h}=0.002905$, and coefficient $c_{h}$ is expressed as  \cite{Boehm.2006}:
\begin{equation}\label{eq:13}
c_{h}=c_{0}+\left[\left(cos\left(\frac{doy-28}{365} \cdot 2\pi +\Psi\right)+1\right)\cdot \frac{c_{11}}{2}+c_{10}\right]\cdot (1-cos \varphi)
\end{equation}
where $c_{0}=0.062$, $doy$ is the day of the year, $\varphi$ is the latitude, $\Psi=0 (\pi)$, $c_{10}=0.000 (0.001)$, and $c_{11}=0.006 (0.006)$ for the Northern (Southern) Hemisphere. For the wet part, the coefficients $b_{w}$ and $c_{w}$ are constants, with values of $b_{w}=0.00146$ and $c_{w}=0.04391$.

Then, the gridded Vienna Mapping Function 1 (VMF1) data are generated from the ECMWF NWM using four global grid files ($2.0^{\circ} \times 2.5^{\circ}$) of $a_{h}$, $a_{w}$, $\Delta L_{h}^{z}$, and $\Delta L_{w}^{z}$. Using VMF1, one can determine the tropospheric delay at any location after 1994 \cite{Kouba.2008,Boehm.2004}. The all VMF1 grid files at any epoch are available at the VMF1 website\footnote{\url{http://ggosatm.hg.tuwien.ac.at/DELAY/GRID/VMFG/}}. In our experiment, the accuracy of the troposphere model correction,  $\Delta t^{tro}_{_{AB}}$, achieves 0.52~ns.

\subsubsection*{Sagnac effect}\label{sec 1.2.4}
The Sagnac effect is caused by the rotation of the Earth.  Concerning a signal propagating from satellite $S$ to the $i$-th station, the Sagnac effect (correction) is expressed as \cite{Tseng.2011}:

\begin{equation}\label{eq:14}
\Delta t^{_{Sac}}_i=- \omega_{e}\frac{x_{s}y_{i}-x_{i}y_{s}}{c^{2}}
\end{equation}
where $\omega_{e}$ is the Earth's rotation rate, which has a relative uncertainty of $\delta \omega_{e}=1.4 \times 10^{-8}$ \cite{Groten.2000}. 

Taking stations A and B into consideration, the corrections of signal delay due to the total Sagnac effect via the CVSTT technique can be expressed as:
\begin{equation}\label{eq:15}
\begin{split}
\Delta t^{_{Sac}}_{_{AB}}&=\Delta t^{_{Sac}}_{_{B}}-\Delta t^{_{Sac}}_{_{A}}\\
&=
[y_{_S}(x_{_B}-x_{_A})-x_{_S}(y_{_B}-y_{_ A})] \omega_{e}/c^{2} \\
\end{split}
\end{equation}

And the accuracy of the Sagnac corrections via the CVSTT technique can be determined as:
\begin{equation}\label{eq:16}
\begin{split}
m^{2}(\Delta t^{_{Sac}}_{_{AB}})&=[(x_{_B}-x_{_A})^{2}m^{2}(\delta y_{_S})+(y_{_B}-y_{_A})^{2}m^{2}(\delta x_{_S}) \\
&+
y_{_S}^{2}m^{2}(\delta x_{_A})+y_{_S}^{2}m^{2}(\delta x_{_B})_+x_{_S}^{2}m^{2}(\delta y_{_A})\\
&+
x_{_S}^{2}m^{2}(\delta y_{_B})](\omega_{e}/c^{2})^{2} \\
&+
[y_{_S}(x_{_B}-x_{_A})-x_{_S}(y_{_B}-y_{_A})] m^{2}(\delta \omega_{e})/c^{2}\ \\
\end{split}
\end{equation}

Taking the BIL and LTS into consideration, the Sagnac effect did not exceed 0.003~ns in the experiment.

\section*{EEMD technique}
\subsection*{EEMD principle}\label{sec 2.1}
Previous studies demonstrate that the ensemble empirical mode decomposition (EEMD) ({\small \url {https://www.worldscientific.com/doi/abs/10.1142/S1793536909000047}}) is an effective technique for isolating target signals from environmental noises \cite{Shen.2014, Wu.2009b}. The EEMD technique is developed from the empirical mode decomposition (EMD) \cite{Huang.1998} ({\small  \url {https://royalsocietypublishing.org/doi/pdf/10.1098/rspa.1998.0193}}). The signals series are decomposed into a series of intrinsic mode functions (IMFs) and a residual trend $r$  after EEMD decomposition. These IMFs series that are sifted stage by stage reflect local characteristics of the signals, while the residual trend $r$ series reflects slow change of the signals. 

\subsection*{Simulation experiments}\label{sec2.2}
\begin{figure}[h!]
	\centering
	\includegraphics[width=1\textwidth]{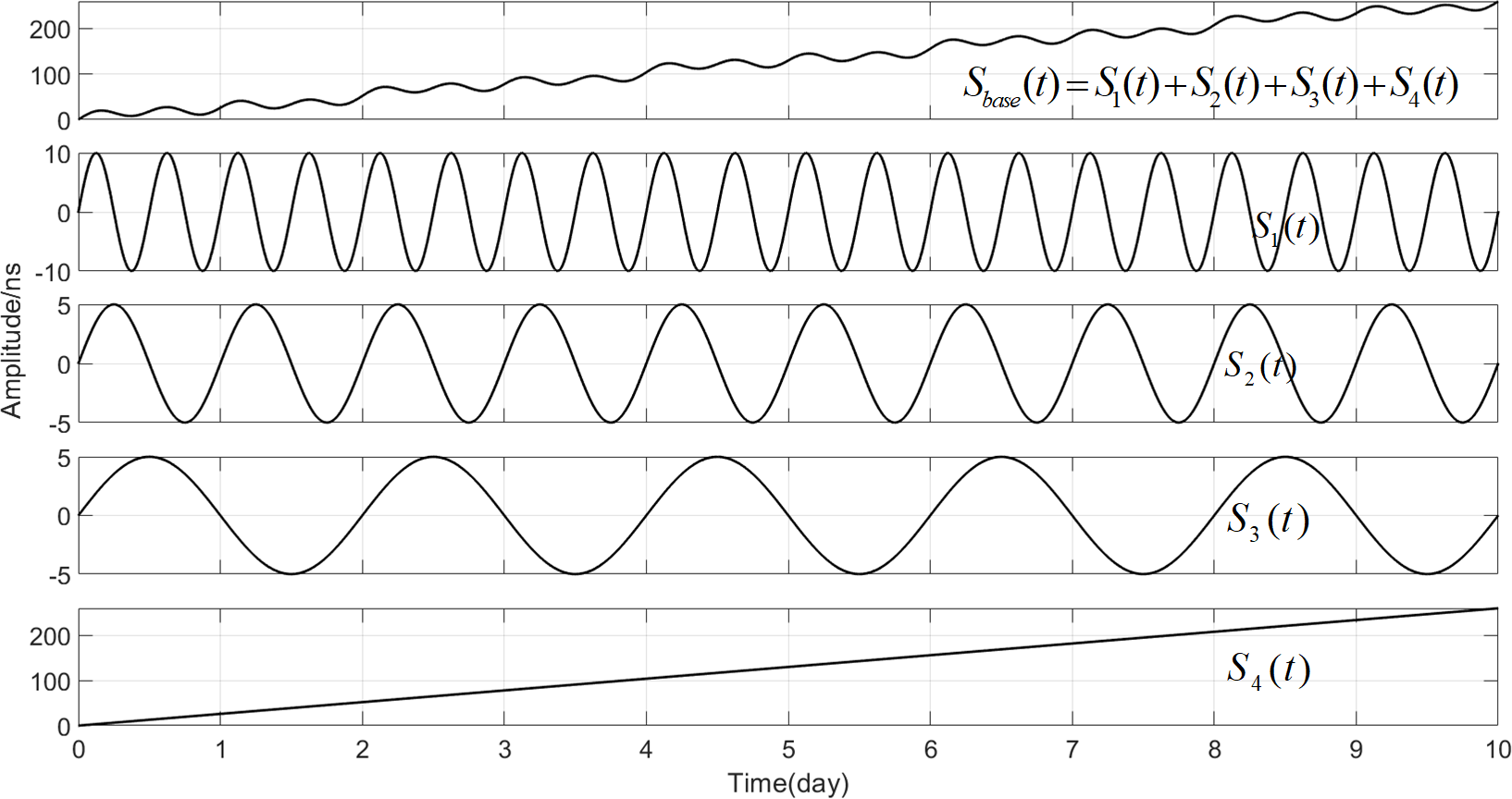}
	\caption{The waveforms of the synthetic series $S_{base}(t)$ (top slot) and different components $S_{i}(t)$ ($i$=1,2,3,4) in the subsequent slots, respectively.}
\label{fig:simulation_original_signals}
\end{figure}
Here, using a simulation experiments we explain the advantages of the EEMD technique. Suppose we have a synthetic series $S_{base}(t)$ that consists of 3 periodic signals series,$S_{i}(t)=A_{i}\cdot \sin(2\pi f_{i}t) \cdot \exp(-10^{-7}t)$(units:ns), where $A_{1}=10$, $A_{2}=A_{3}=5 $, $f_{1}=2/86400$ Hz, $f_{2}=1/86400$ Hz, $f_{3}=0.5/86400$ Hz, respectively, and a linear signal series, $S_{4}(t)= 3 \cdot 10^{-4}t-10^{-7}$, with a data length of 10 days and sampling interval 960 seconds. The $S_{base}(t)$ can be expressed as:
\begin{equation}\label{eq:17}
\begin{split}
S_{base}(t)&=S_{1}(t)+S_{2}(t)+S_{3}(t)+S_{4}(t) \\
&=
10\cdot \sin(2\pi\cdot\dfrac{2}{86400}\cdot t) \cdot \exp(-10^{-7}t)\\
&+5\cdot \sin(2\pi\cdot\dfrac{1}{86400}\cdot t) \cdot \exp(-10^{-7}t) \\
&+5\cdot \sin(2\pi\cdot\dfrac{0.5}{86400}\cdot t) \cdot \exp(-10^{-7}t)\\ 
&+(3 \cdot 10^{-4}\cdot t-10^{-7}) \\
\end{split}
\end{equation}

\begin{figure}[h!]
	\centering
	\includegraphics[width=1\textwidth]{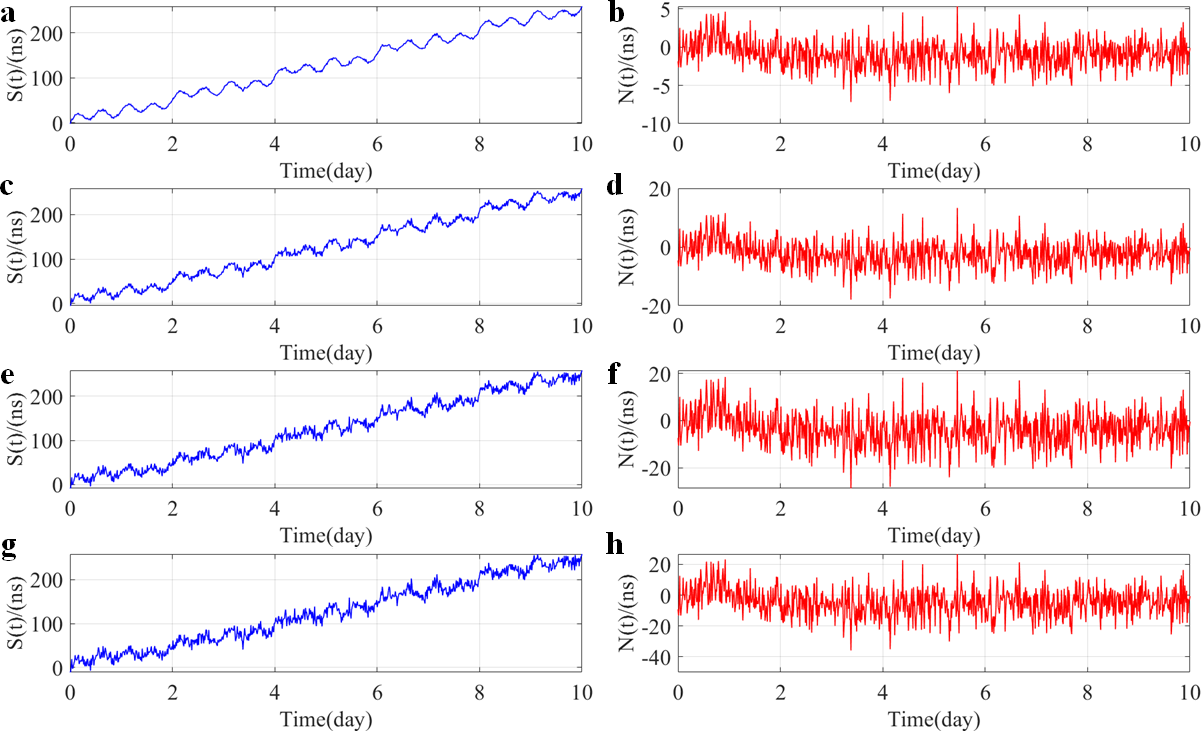}
	\caption{The waveforms of the constructed signal series $S(t)$ and different magnitudes of the noise series $N(t)$. (a), (c), (e) and (g) are the constructed signal series $S(t)$ which are the sum of $S_{base}(t)$ and corresponding $N(t)$ with magnitude of 2$\%$, 5$\%$, 8$\%$ and 10$\%$ of $S_{base}(t)$, respectively; (b), (d), (f) and (h) are the noise series $N(t)$ with magnitudes of  2$\%$, 5$\%$, 8$\%$ and 10$\%$ of $S_{base}(t)$, respectively. }
\label{fig:simulation_S_t__and_N_t_}
\end{figure}

The results of the constructed series are shown in  Fig. \ref{fig:simulation_original_signals}. We added noise signals $N(t)$ into $S_{base}(t)$ with 2$\%$ (Case 1), 5$\%$ (Case 2), 8$\%$(Case 3) and 10$\%$ (Case 4) of the standard deviation (STD) of $S_{base}(t)$, and then construct a signal series $S(t)$ which contains noise. The $S(t)$ can be expressed as:
\begin{equation}\label{eq:18}
S(t)=S_{base}(t)+N(t)
\end{equation}
where the $N(t)$ is a noise series, consisting of five types of noises, expressed as \cite{Li.2011,Zhai.2012,Ashby.2015}:
\begin{equation}\label{eq:19}
\begin{split}
N(t)&=N_{W-PM}(t)+N_{F-PM}(t)+N_{W-FM}(t)\\
&+
N_{F-FM}(t)+N_{RW-FM}(t)\\
\end{split}
\end{equation}
where $N_{W-PM}(t)$ is the white noise phase modulation (W-PM), $N_{F-PM}(t)$ is the flicker noise phase modulation (F-PM), $N_{W-FM}(t)$ is the white noise frequency modulation (W-FM), $N_{F-FM}(t)$ is the flicker noise frequency modulation (F-FM), $N_{RW-FM}(t)$ is the random walk noise frequency modulation (RW-FM).

\begin{figure}
\centering
\subfigure[]{
\includegraphics[width=0.46\textwidth]{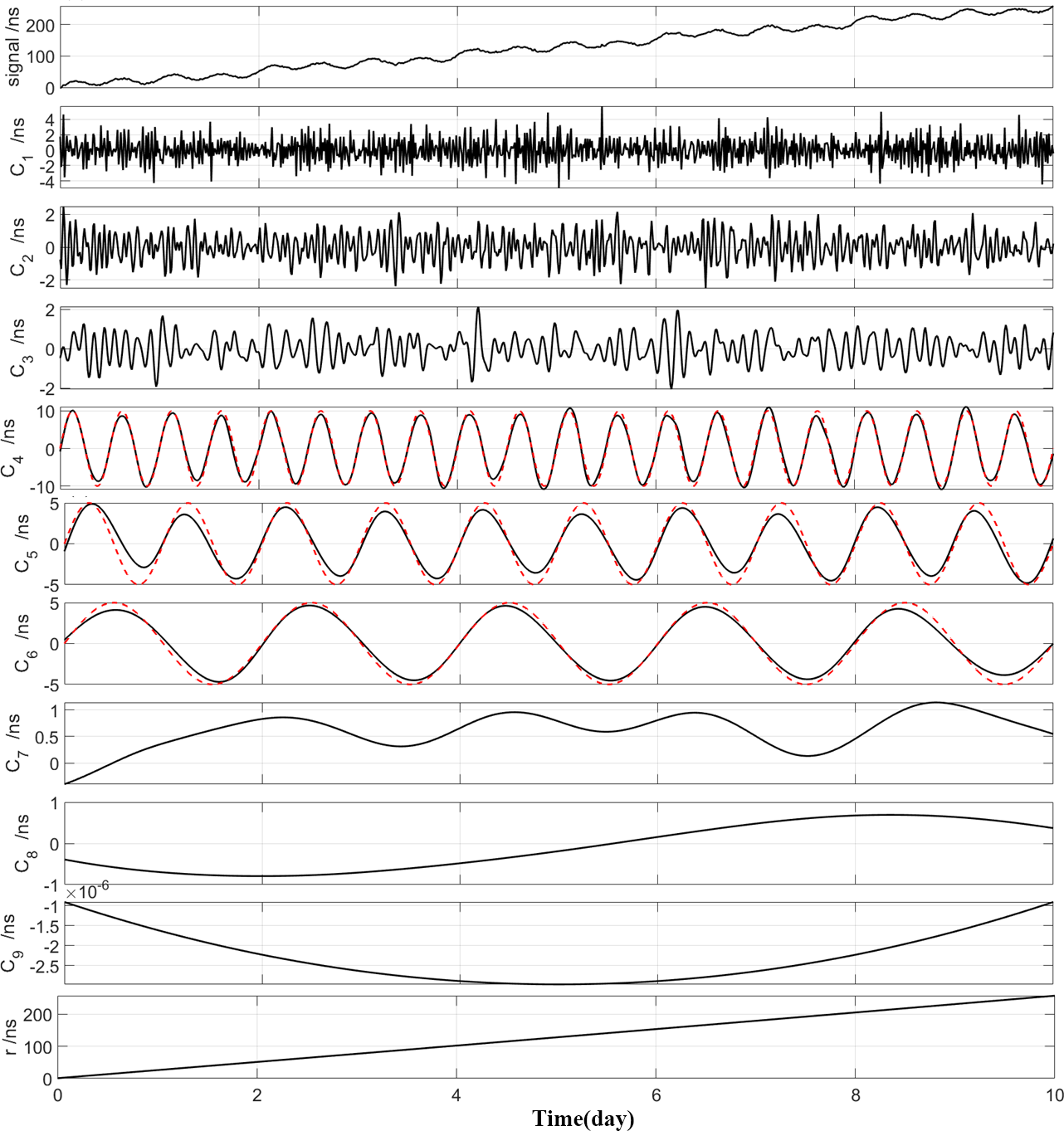}
}
\quad
\subfigure[]{
\includegraphics[width=0.46\textwidth]{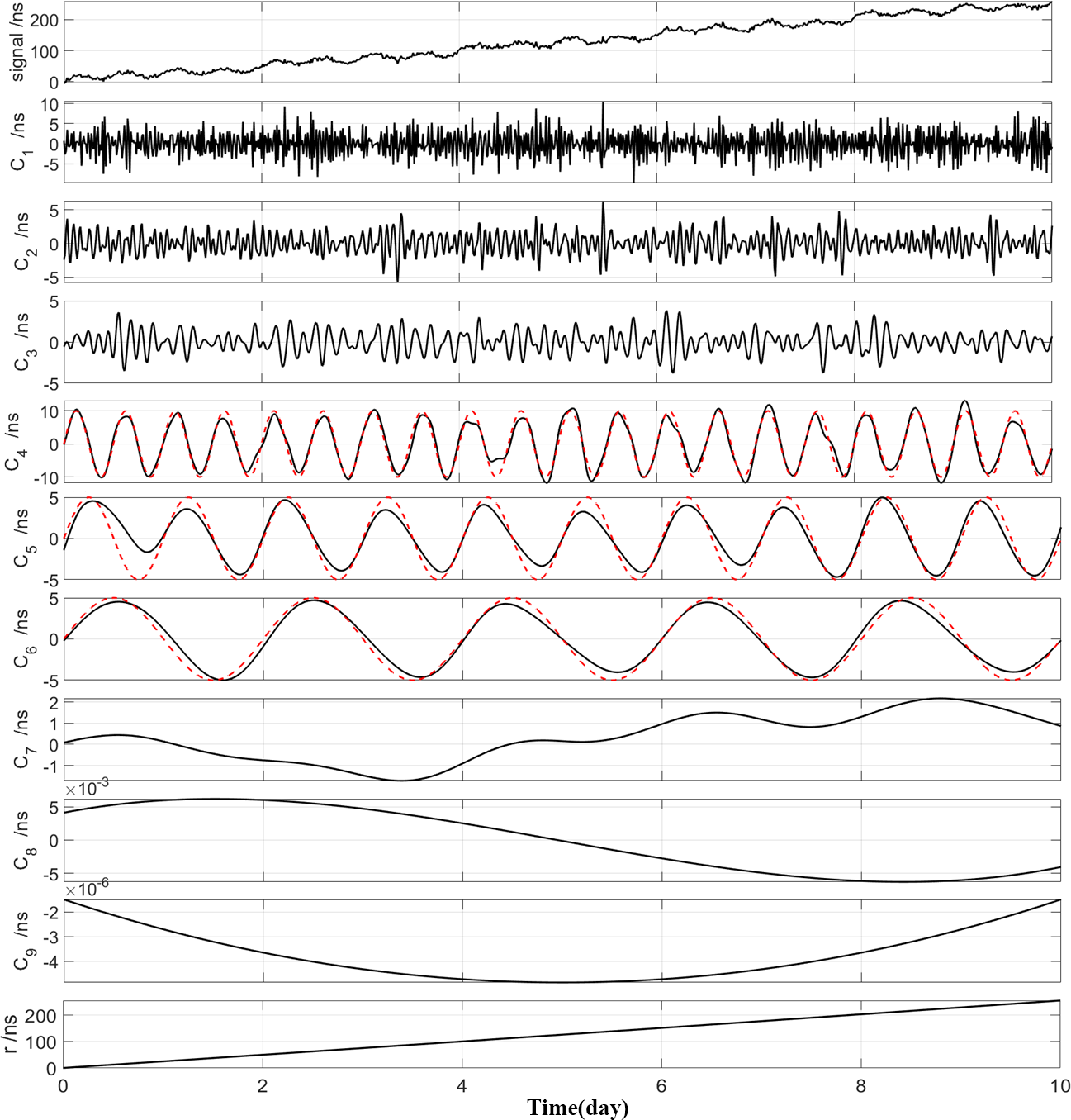}
}
\quad
\subfigure[]{
\includegraphics[width=0.46\textwidth]{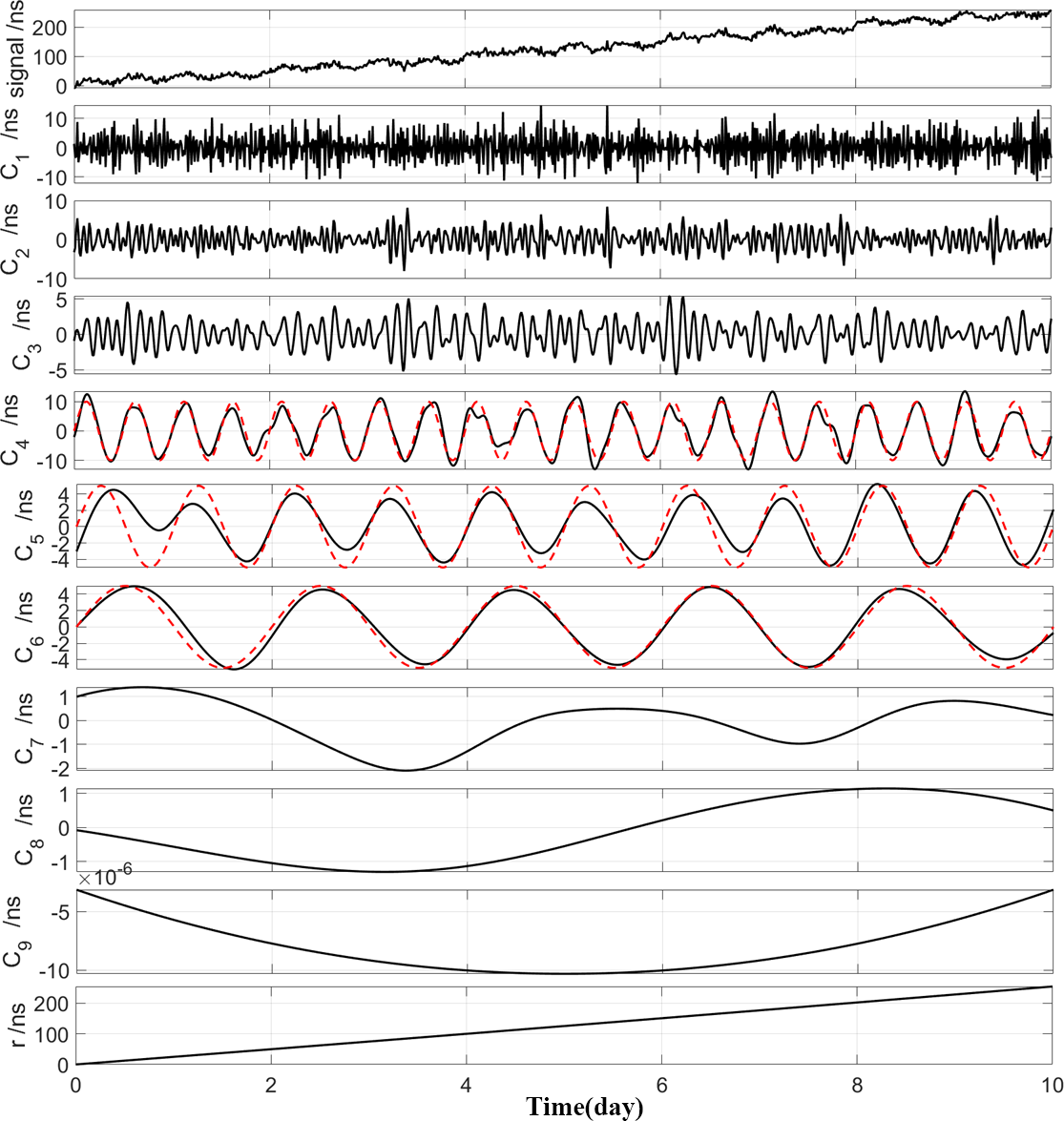}
}
\quad
\subfigure[]{
\includegraphics[width=0.46\textwidth]{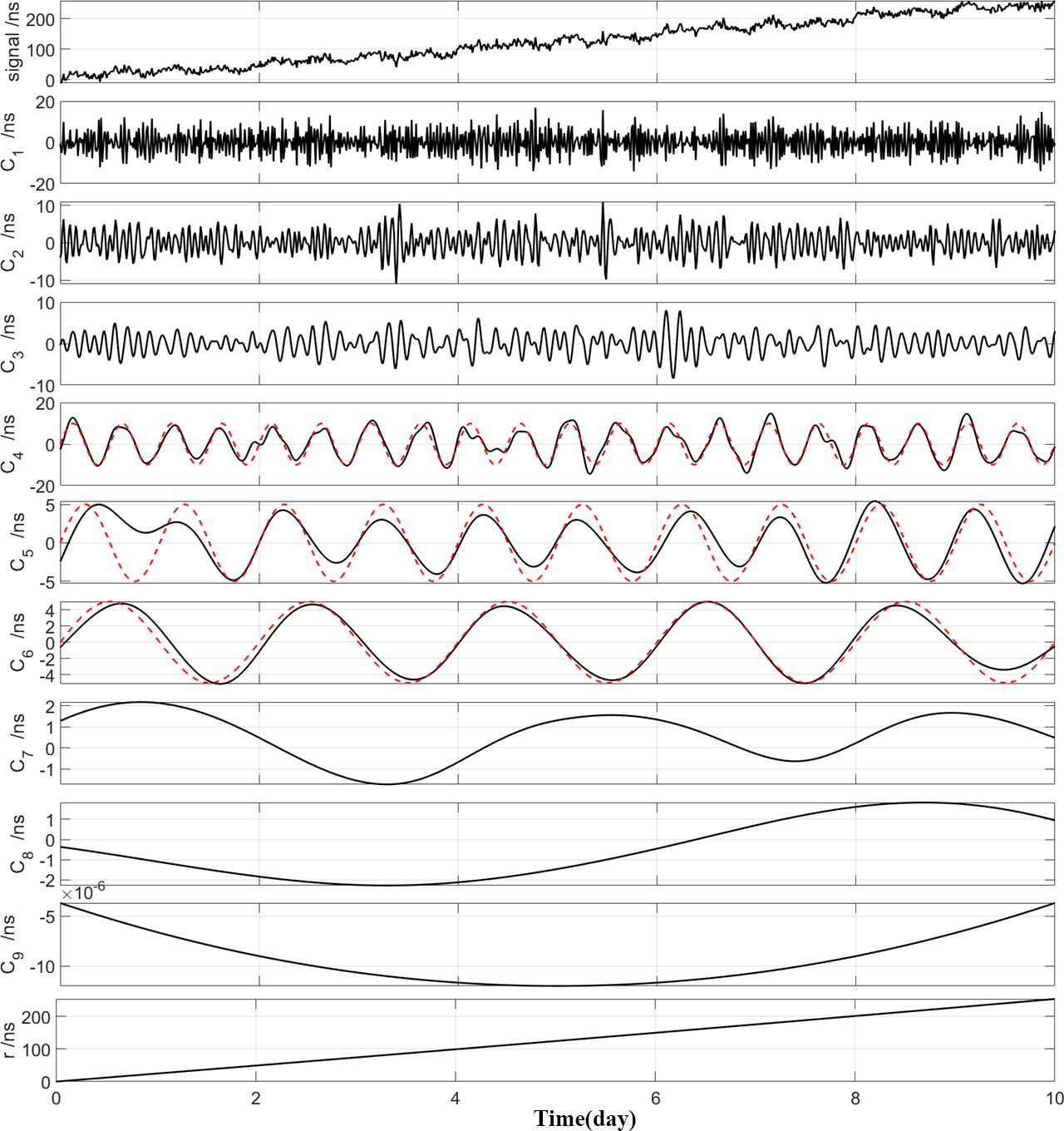}
}
\caption{The resulting IMFs from the constructed signal series $S(t)$ in (a) case~1, (b) case~2, (c) case~3 and (d) case~4, respectively. The constructed signal series $S(t)$, IMFs and the residual trend r are denoted with black curves; the periodic signals $S_{1}(t)$, $S_{2}(t)$ and $S_{3}(t)$ are denoted with red curves.}
\label{fig:cases_imfs}
\end{figure}

The constructed signal series $S(t)$ with different noise magnitudes are shown in Fig. \ref{fig:simulation_S_t__and_N_t_}. For our present purpose, the linear signal series $S_{4}(t)$ is the target signal which need to be extracted from the synthetic signals series $S(t)$. The EEMD technique is applied to identify the periodic signals $S_{1}(t)$, $S_{2}(t)$ and $S_{3}(t)$ from $S(t)$. After EEMD decomposition, we use the index IO to check completeness of the decomposition. In our simulation experiments, the values of IO are 0.0015 (Case 1), 0.0017 (Case 2), 0.0016 (Case 3) and 0.0017 (Case 4), respectively, suggesting that signal series $S(t)$ are effectively decomposed in four different cases. The decomposed IMFs are shown in Fig. \ref{fig:cases_imfs}, and we found that the periodic signals $S_{1}(t)$, $S_{2}(t)$ and $S_{3}(t)$ can be clearly identified, respectively. The red dotted curves in Fig. \ref{fig:cases_imfs} denote the original signals $S_{i}(t)$ ($i$ = 1, 2, 3) for comparison purpose. 

\begin{figure}[h!]
	\centering
	\includegraphics[width=0.8\textwidth]{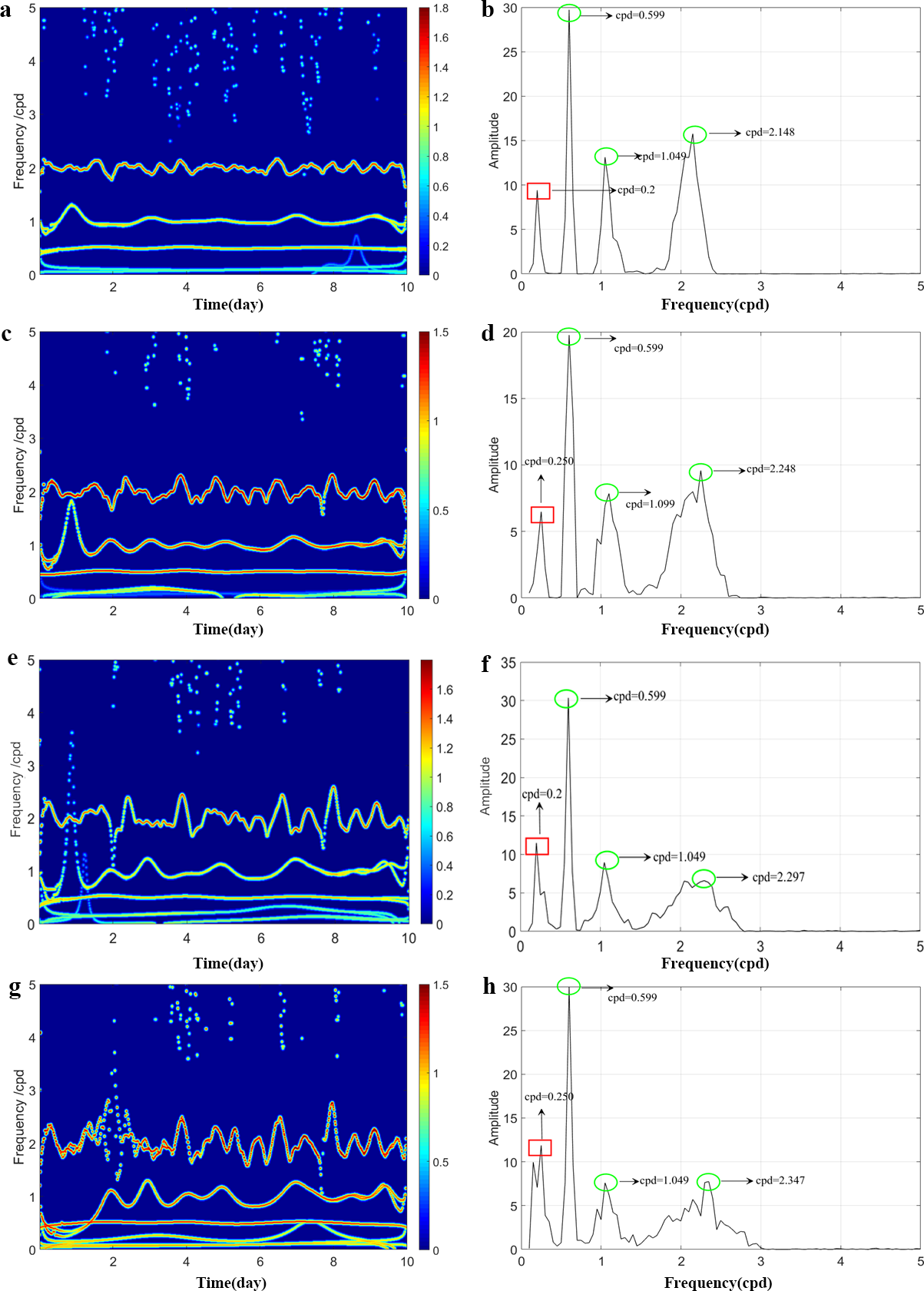}
	\caption{The frequency variations of each IMF (expect for trend r) decomposed from S(t) in Case 1((a) and (b)), Case 2 ((c) and (d)), Case 3 ((e) and (f)) and Case 4 ((g) and (h)). (a), (c), (e) and (g) are the Hilbert spectrum of IMFs, where the color variations occurred in each skeleton curve represent the corresponding energy variations of each IMF;  (b), (d), (f) and (h) are the corresponding marginal spectrum of IMFs.}
	\label{fig:cases_HHT}
\end{figure}

Further, the Hilbert transform (HT) is executed to display the variety of instantaneous frequencies for each IMF component. The corresponding Hilbert spectra and marginal spectra for IMFs (expect for r) in three cases are shown in  Fig. \ref{fig:cases_HHT}, where the subfigures (a), (c), (e) and (g) show the frequency variations of each IMF in four cases, and the subfigures (b), (d), (f) and (h) show measures of the total amplitude (or energy) contribution from each frequency value in four cases, representing the cumulated amplitude over the entire data span in a probabilistic sense \cite{Huang.1998}.  

From Fig. \ref{fig:cases_HHT}, we see that the mode-mixing problem exists in EEMD decomposition, and it becomes more obvious as noise increases from 2$\%$ to 10$\%$. However, the three periodic signals $S_{1}(t)$, $S_{2}(t)$, and $S_{3}(t)$ can be identified clearly. The set frequencies of $S_{1}(t)$, $S_{2}(t)$, and $S_{3}(t)$ are 2 circle per day (cpd), 1 cpd and 0.5 cpd, respectively, denoted as $f_{1-set}=2$ cpd, $f_{2-set}=1$ cpd, and $f_{3-set}=0.5$ cpd. After EEMD decomposition, the marginal spectra show that the corresponding values are $f_{1-case1}=2.148$ cpd, $f_{2-case1}=1.049$ cpd, $f_{3-case1}=0.599$ cpd (Case 1); $f_{1-case2}=2.248$ cpd, $f_{2-case2}=1.099$ cpd, $f_{3-case2}=0.599$ cpd (Case 2); $f_{1-case3}=2.297$ cpd, $f_{2-case3}=1.049$ cpd, $f_{3-case3}=0.599$ cpd (Case 3); $f_{1-case4}=2.347$ cpd, $f_{2-case4}=1.049$ cpd, $f_{3-case4}=0.599$ cpd (Case 4), respectively. And the detected signals corresponding to the original set signals $S_{1}(t)$, $S_{2}(t)$ and $S_{3}(t)$ are shown by the peaks denoted in green circles. In addition, there appear not-real signals with frequencies around 0.2 cpd after EEMD decomposition, which are denoted as $F(t)$ and shown by the peaks denoted in red rectangle. The  $F(t)$  might be meaningless in physical explaining, and it will be an interference if we focus on periodic signals. In this study, however, the target is to detect and identify the linear signal $S_{4}(t)$, which means that all periodic signals are useless and should be removed. After all these periodic signals are removed, we reconstruct a new signal series by summing the residual IMFs and the residual trend r. The reconstructed signal series is denoted as $S^{'}(t)$.

\begin{figure}[h!]
	\centering
	\includegraphics[width=1\textwidth]{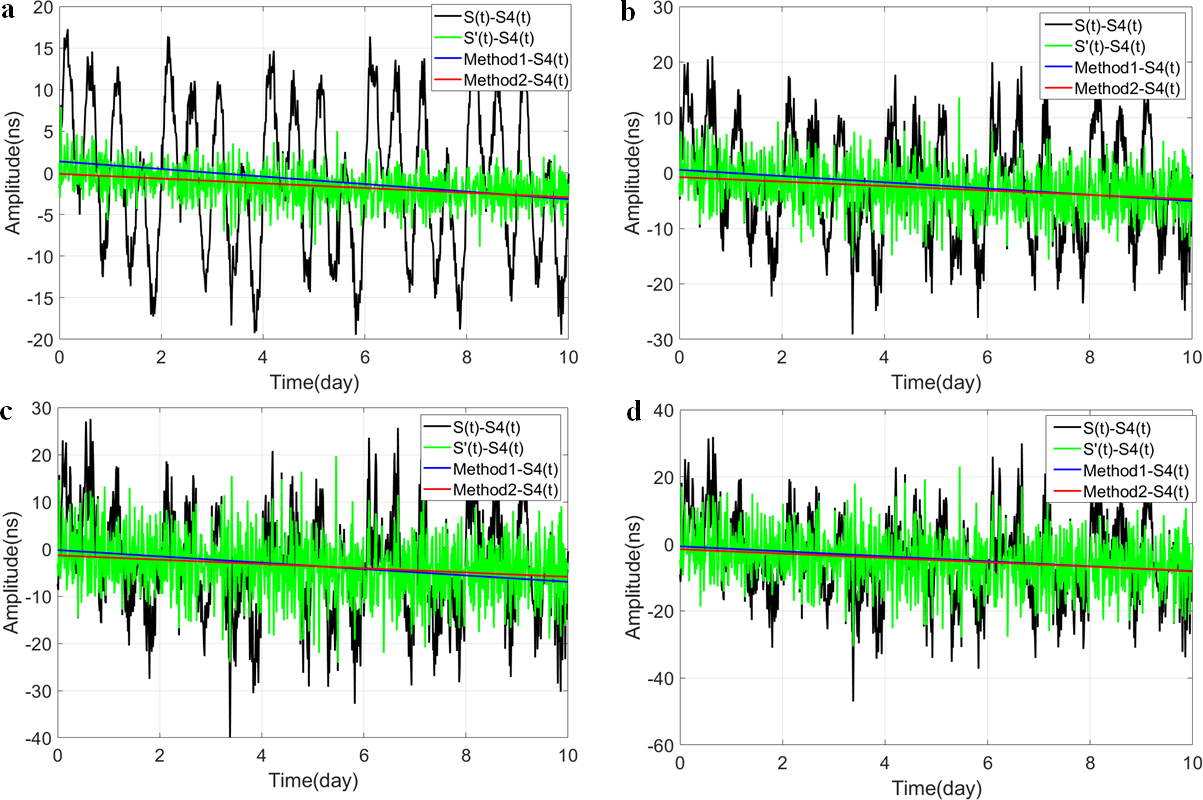}
	\caption{The comparison of the linear fitting for the signal series $S(t)$ (Method1) and the linear fitting for the reconstructed signal series $S^{'}(t)$ (Method2). $S_{4}(t)$ is a given real signal series. (a), (b) (c) and (d) denote the cases of noise magnitudes of 2$\%$, 5$\%$, 8$\%$ and 10$\%$ of $S_{base}(t)$, respectively.}
	\label{fig:cases_comparison}
\end{figure}

We take the signal series $S_{4}(t)$ as real signal series, and try to recovery it by two different methods. The first method is to perform a least squares linear fitting on the signal series $S(t)$  directly; and the second method is to perform EEMD decomposition on $S(t)$ and then reconstruct the new signal series $S^{'}(t)$ by removing periodic series IMFs, and the least squares linear fitting is performed on $S^{'}(t)$. The results are shown in Fig. \ref{fig:cases_comparison}.

Here we use the standard deviation (STD) of the difference series between the real signal series $S_{4}(t)$ and that determined by Method 1 or 2 to evaluate the reliability of the two methods. Using Method 1, namely the direct least squared linear fitting of the series $S(t)$, the STDs of the results are 1.31 ns (Case 1), 1.63 ns (Case 2), 1.94 ns (Case 3) and 2.16 ns (Case 4), respectively. Using Method 2, namely the least squared linear fitting of the reconstructed series $S^{'}(t)$, which is obtained after removing the periodic series via EEMD technique, the STDs of the results are 0.82 ns (Case 1), 1.16 ns (Case 2), 1.31 ns (Case 3) and 1.87 ns (Case 4), respectively. The comparative results clearly suggest that the EEMD technique is effective for extracting the linear signals of interest by a priori removing the contaminated periodic signals from the original observations.

\section*{Experimental results}\label{sec：3}
\begin{figure*}[h!]
	\centering
    \includegraphics[width=1\textwidth]{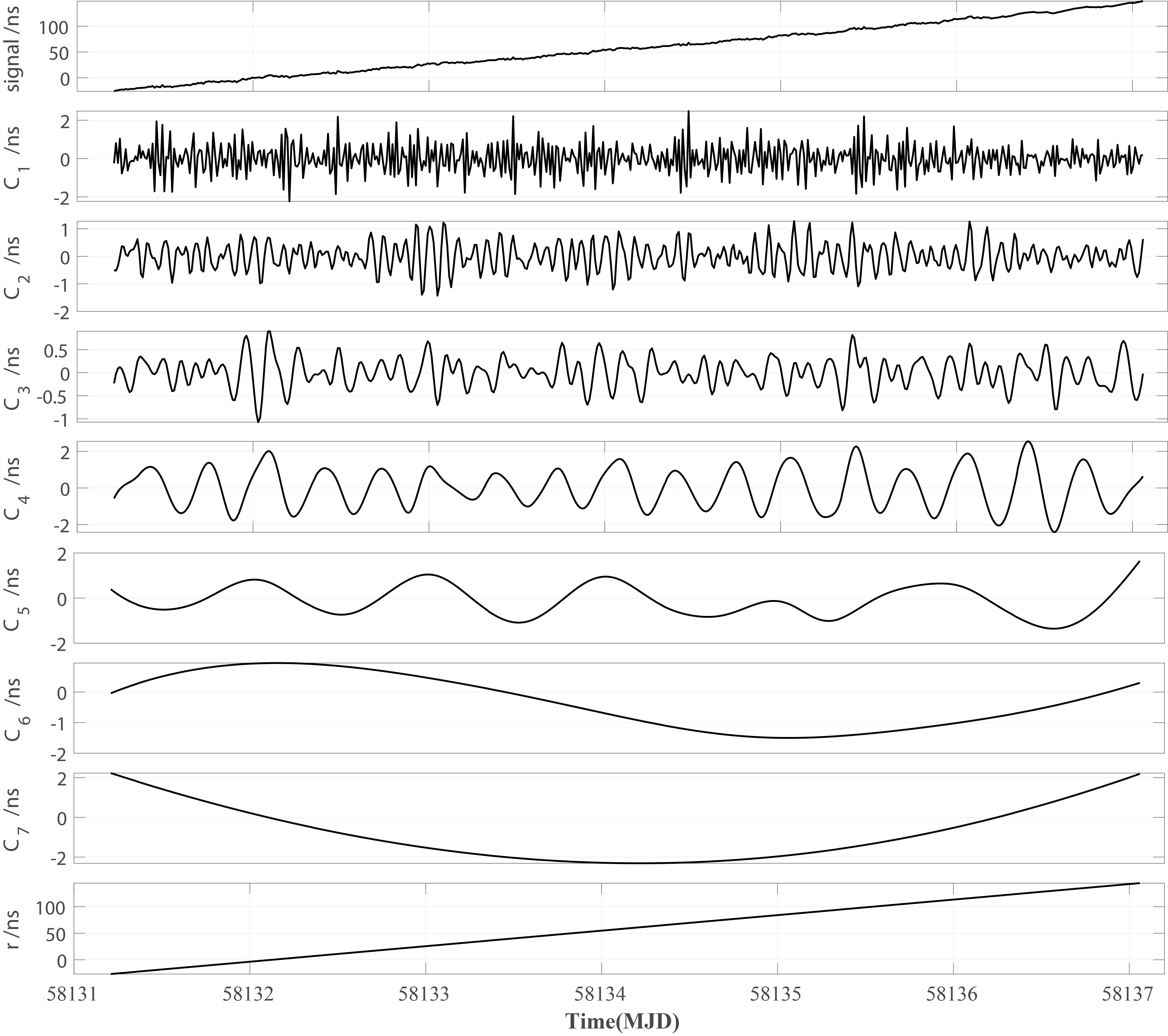}
	\caption{The resulting IMFs from the time difference series of preprocessed data set, $\Delta t_{AB}(t)$, in Period~1 with both $C_{A}$ and $C_{B}$ located at BIL, which lasted from MJD 58131 to 58137.}
	\label{fig:Period1_IMFs}
\end{figure*}

\begin{figure*}[h!]
\centering
\includegraphics[width=1\textwidth]{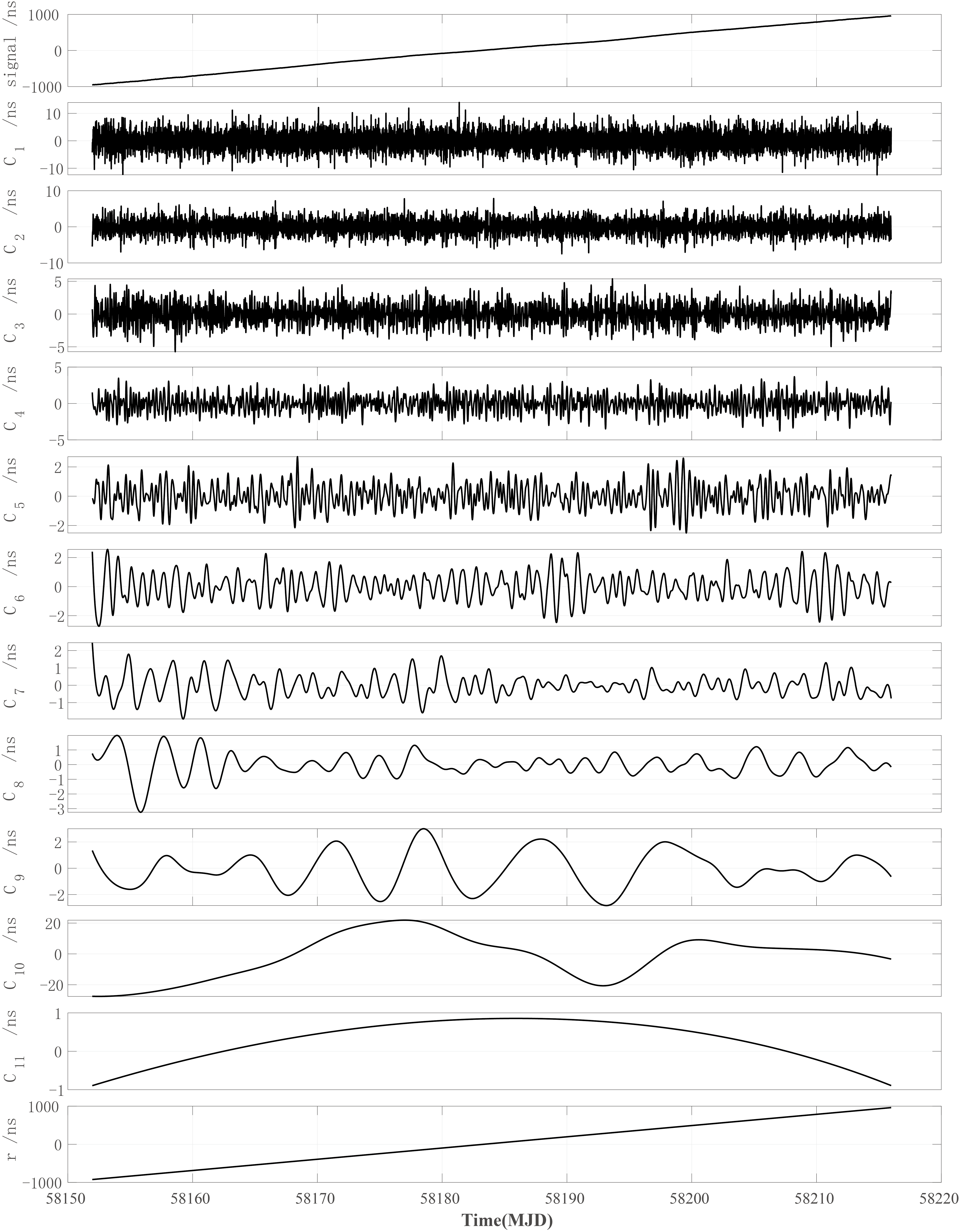}
\caption{The resulting IMFs from the time difference series of preprocessed data set, $\Delta t_{AB}(t)$, in Period~2, with $C_{A}$ located at BIL and $C_{B}$ located at LTS, which lasted from MJD 58150 to 58215.}
\label{fig:Period2_IMFs}
\end{figure*}

\begin{figure*}[h!]
\centering
\includegraphics[width=1\textwidth]{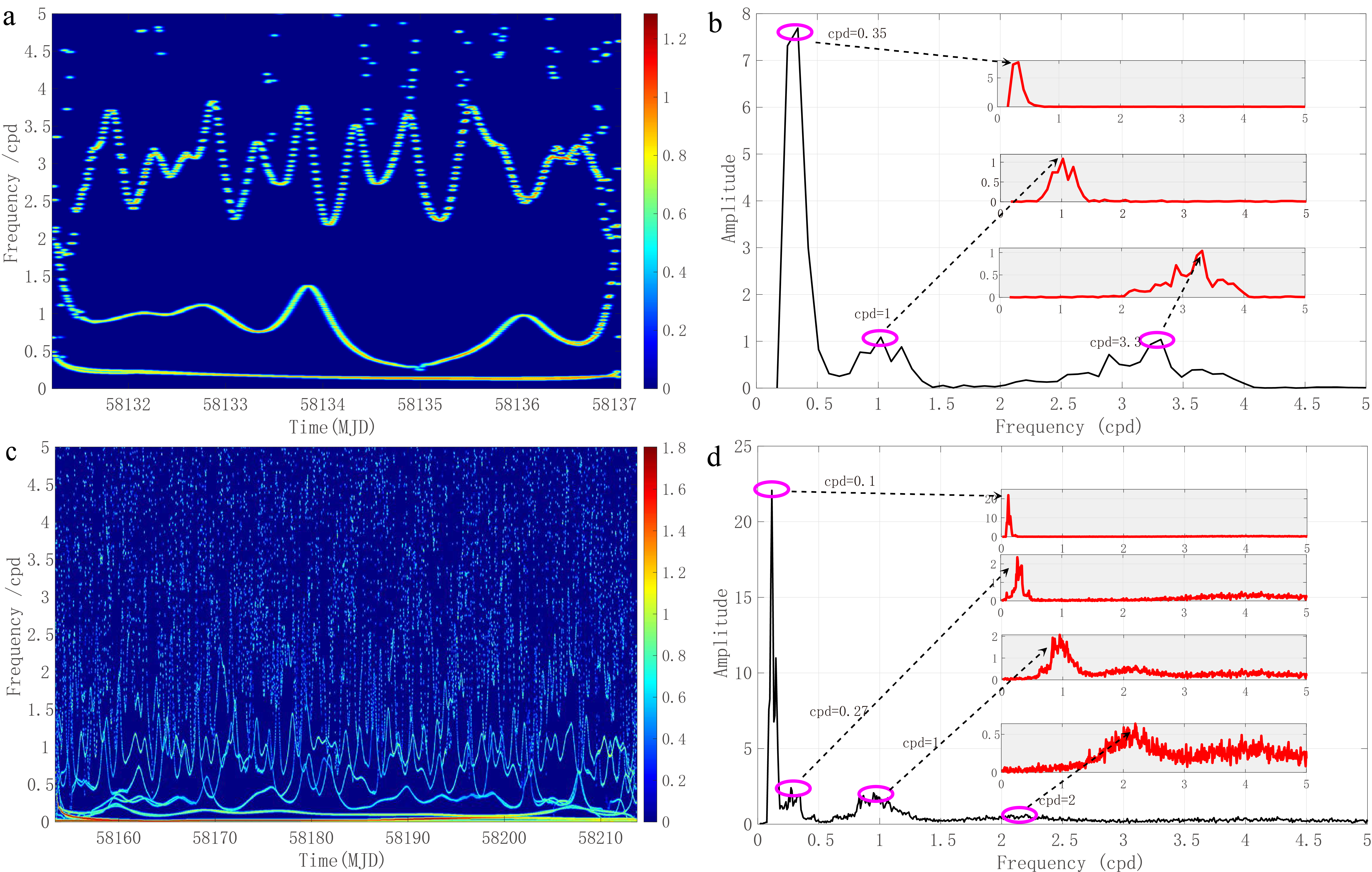}
\caption{The Hilbert spectra and marginal spectra of EEMD decomposition in Period~1 and Period~2. (a) and (b) are the Hilbert spectra and marginal spectra in Period~1, respectively. (c) and (d) are the Hilbert spectra and marginal spectra in Period~2, respectively. In Period~1, the periodic signals with frequencies around 0.3~cpd, 1cpd and 3.3~cpd are separated and  then removed. In Period~2, the periodic signals with frequencies around 0.1~cpd, 0.3~cpd, 1cpd and 2~cpd are separated and then removed.}
\label{fig:HHTAndMar}
\end{figure*}

By applying EEMD technique, the time difference series of preprocessed data sets, $\Delta t_{AB}(t)$, in Period 1 and Period 2
are decomposed into a series of intrinsic mode functions and a long trend component, r, respectively. And the corresponding decompisiotn are shown in Fig. \ref{fig:Period1_IMFs} and Fig. \ref{fig:Period2_IMFs}. And the corresponding Hilbert spectra and marginal spectra in Period 1 and Period 2 are determined, which are given in Fig. \ref{fig:HHTAndMar}, respectively.The marginal spectra which  represent the cumulated amplitude over the entire data span in a probabilistic sense gives the energy contributions from each frequency value. From Fig. \ref{fig:HHTAndMar}, the periodic signals with frequencies of around 0.35 cpd, 1 cpd and 3.3 cpd of zero-baseline measurement are detected; and the periodic signals with frequencies of around 0.1 cpd, 0.3 cpd, 1 cpd and 2 cpd of geopotential difference measurement are detected. 

\begin{figure}[h!]
	\centering
	\includegraphics[width=1\textwidth]{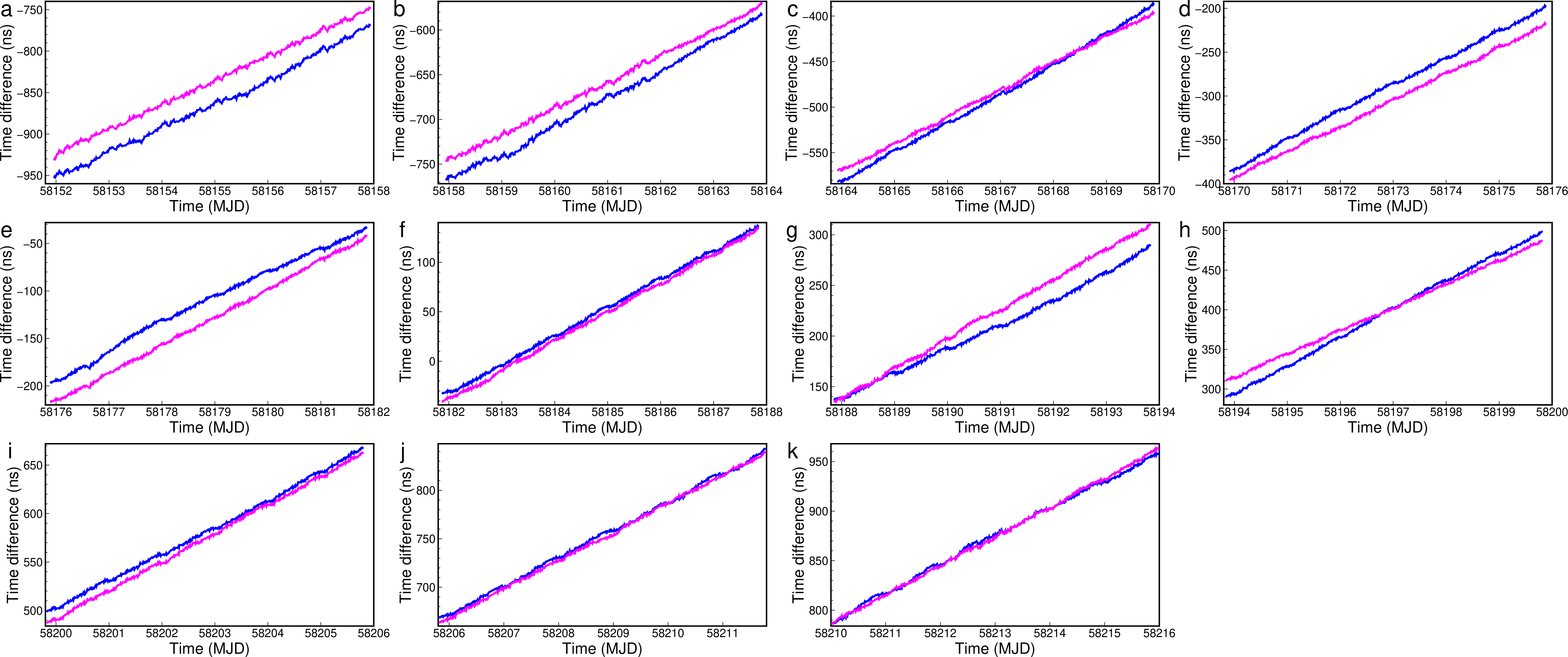}
	\caption{The time difference series of each segment in Period 2 after the implementation of grouping strategy with 6-days measurement as diferent segments. The blue curves are the preprocessed data sets, $\Delta t_{_{AB}}(t)$; the purple curves are the corresponding after-EEMD data sets, $\Delta t_{_{AB}}^{re}(t)$.}
	\label{fig:Period2_Groups_clockoffsets}
\end{figure}

\begin{figure}[h!]
	\centering
	\includegraphics[width=1\textwidth]{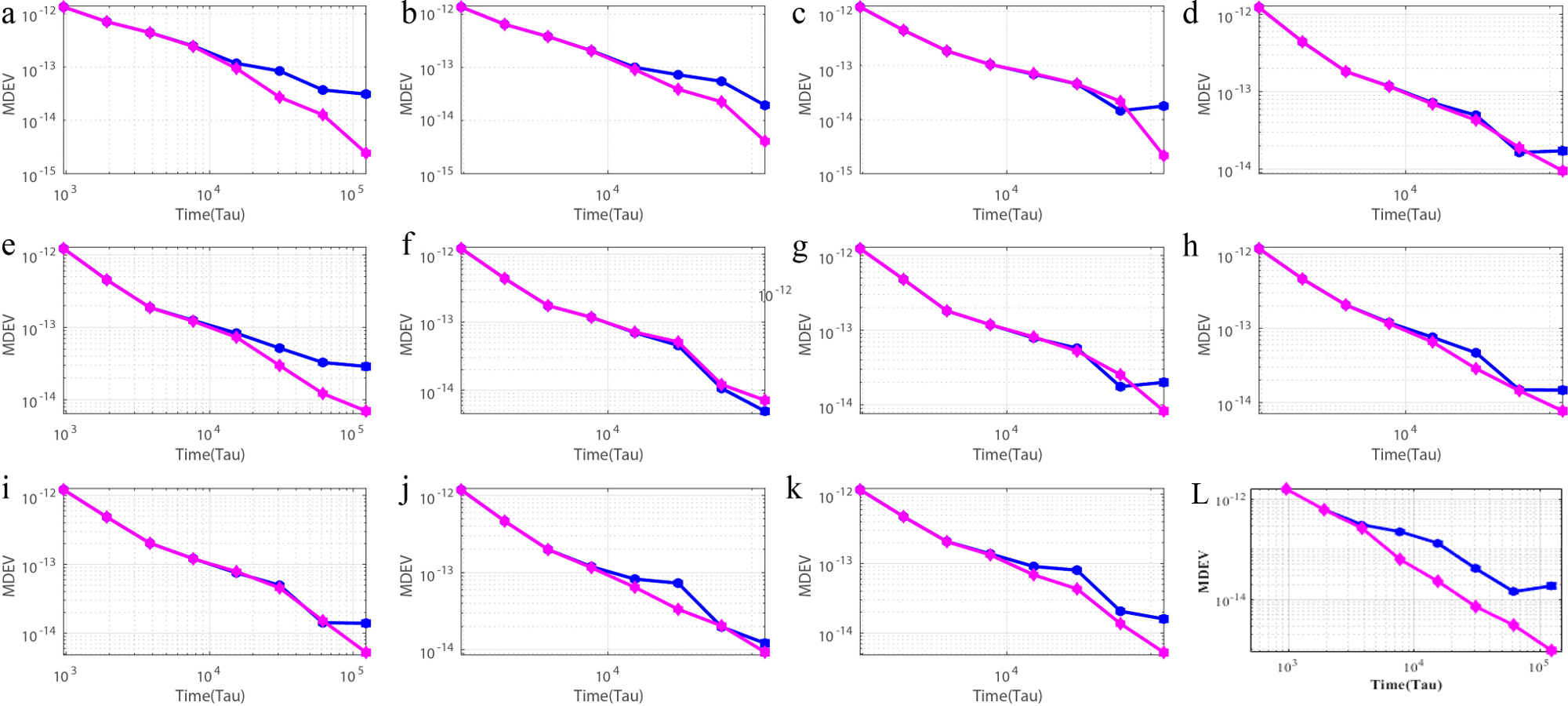}
	\caption{The relative frequency stabilities of the two clocks in Period 1 and Period 2 in the experiment. The blue and purple curves denote the stabilities of the data sets before and after EEMD filtering, respectively. The subfigures from (a) to (k) are the MDEVs of each segment in Period 2, the subfigure (L) is the MDEV of Period 1 as comparison.}
	\label{fig:Period1AndPeriod2MDEV}
\end{figure}

After EEMD technique, the reconstruct time difference series, $\Delta t_{_{AB}}^{re}(t)$ and corresponding  MDEVs of each segment in Period 2 are given in Fig. \ref{fig:Period2_Groups_clockoffsets} and Fig. \ref{fig:Period1AndPeriod2MDEV}, respectively.

\end{document}